\documentclass{ws_tasi}
\input axodraw.sty
\usepackage{epsf}

\begin{document}

\def\cal{\it}
\def\beq{\begin{equation}}
\def\eeq{\end{equation}}
\def\eq{\beq\eeq}
\def\beqn{\begin{eqnarray}}
\def\eeqn{\end{eqnarray}}
\def\ra{\rightarrow}

\input rotate
\def\flipright#1{%
   \setbox0=\vbox{\epsfbox{#1}}
   \setbox1=\vbox{\rotr0}
   \centerline{\copy1}}                 
\def\flipleft#1{%
   \setbox0=\vbox{\epsfbox{#1}}
   \setbox1=\vbox{\rotl0}
   \centerline{\copy1}}                 

\title{
The Top Quark, QCD, and New Physics
 }
\author{S.~Dawson\footnote{\uppercase{T}his work is supported by
the \uppercase{U.S. D}epartment 
of \uppercase{E}nergy under contract number 
\uppercase{DE-AC}02-76-\uppercase{CH}-00016.
}}

\address
{ Physics Department, Brookhaven National Laboratory,\\
Upton, NY 11973, USA\\ 
\texttt{dawson@bnl.gov}}

\maketitle

\abstracts{ 
The role of the top quark in completing the Standard Model quark sector
 is reviewed, along with a discussion of
production, decay, and theoretical restrictions
on the top quark properties.  
Particular attention is paid to the top quark as a laboratory for perturbative 
QCD.  As examples of the relevance of QCD corrections in
the top quark sector,  the
 calculation of $e^+e^-\rightarrow t {\overline t}$ at next-
to-leading-order  QCD using the phase space slicing algorithm and the
implications of a precision measurement of the top quark mass 
 are  discussed in detail.
The associated production of a $t {\overline t}$ pair
and a Higgs boson in either $e^+e^-$ or hadronic collisions is presented
at next-to-leading-order
 QCD and its importance for a measurement of the 
top quark Yukawa coupling emphasized.
 Implications
of the heavy top quark mass for model builders are 
briefly examined, with the minimal
supersymmetric Standard Model and topcolor discussed
 as specific examples.}
\section{Introduction}
\label{sec:intro}

Long before its discovery in 1995,\cite{tdis}
 the top quark was regarded as an
essential ingredient of the Standard Model (SM) of particle physics.  Its
existence, and many of its properties, are determined by requiring
the consistency of the Standard Model.  
These requirements (discussed in Section 2)
 specify the couplings of the top quark to the
$SU(3)\times SU(2)_L \times 
U(1)_Y$ gauge bosons and many of the
top quark properties:
\begin{itemize}
\item
$Q_{em}^t={2\over 3 }\mid e \mid$
\item
Weak isospin partner of $b$ quark: $T_3^t={1\over 2}$
\item
Color triplet
\item
spin-${1\over 2}$
\end{itemize}
The top quark mass, which is measured  at CDF and D0 to
be $M_t=174.3\pm 5.1~GeV$,\cite{tmass} is not predicted in the Standard
Model, but
 is restricted by precision electroweak
measurements.\cite{ewtop,pdg}   In order to confirm that the observed top
quark is that predicted by the Standard Model, all of the properties
listed above must be experimentally verified by direct observation.
Section 3 contains a discussion of the measurements of top quark
properties at the Tevatron and surveys the improvements expected at future
colliders. Recent reviews of top quark physics can be found in Ref.
\refcite{toprevs}.

In Section 4, we
 discuss the top  quark as a laboratory for perturbative QCD.  At the top
quark mass scale, the strong coupling constant is small,
 $\alpha_s(M_t)\sim 0.1$, and so QCD effects involving
the top quark are well behaved and we expect a perturbation
series in $\alpha_s$ to converge rapidly.  The process
$e^+e^-\rightarrow t {\overline t}$ provides  an
example of QCD effects in top quark production
and the next-to-leading-order (NLO) QCD corrections are described
in detail using the phase space slicing (PSS)
algorithm.  A knowledge of these
higher order QCD 
corrections is vital for extracting a precise value of the top quark
mass from  the threshold behavior of the
$e^+e^-\rightarrow t {\overline t}$ cross section.  As a further example
of the role of 
 QCD effects in the top quark sector, we consider the associated
production of $t {\overline t}h$ and the implications for measuring the
top quark Yukawa coupling.

Finally, in Section 5, we discuss the importance of the
top  quark mass  in model building.
Since the top quark is heavy,
  it  is expected to play a special role in
elucidating the source of fermion masses.
We begin by discussing the significance of the top
quark in supersymmetric models, and finish with a brief discussion 
of the top quark and models with dynamical symmetry breaking.

\section{Who needs a top quark?}

In this section we discuss some of the reasons why the 
top quark was believed to exist 
even before its experimental discovery. 
These considerations
 fall into three general categories:  theoretical consistency
of the Standard Model gauge theory (anomaly cancellation), consistency
of $b$ quark measurements with SM predictions,
 and consistency of precision measurements with the SM.
 We then turn to a discussion
of top quark production and decay mechanisms at the Tevatron and
the LHC.

The particles of the first generation of fermions, along with the Higgs
doublet, are shown in Table \ref{smpart}, with their gauge quantum numbers.
The third generation is assumed to follow the same pattern, with the 
left-handed top and bottom quarks forming an $SU(2)_L$ doublet with 
hypercharge, $Y={1\over 6}$.  The right-handed top and bottom quarks
are $SU(2)_L$ singlets.  Our normalization is such that:
\begin{equation}
Q_{em}=T_3+Y
\end{equation}
with $T_3=\pm{1\over 2}$.

\begin{table}[ph]
\tbl{Fermions in the first 
generation of the  Standard Model
 and their $SU(3)\times SU(2)_L\times
U(1)_Y$ quantum numbers. }
{\footnotesize
\begin{tabular}{|c|c|c|r|}
\hline
Field & $SU(3)$& $SU(2)_L$& $U(1)_Y$\\
\hline\hline 
($u_L,d_L$)   &    $3$          & $2$&  $~{1\over 6}$
\\
$u_R$, & $3$ & $1$& $~{2\over 3}$
\\
$d_R$, & $3$ & $1$&  $-{1\over 3}$
\\
$(\nu_L,e_L)$   & $1$         & $2$& $~-{1\over 2}$
\\
$e_R$ & $1$             & $1$& $-1$ 
\\
${ H}$ &$1$    
         & $2$& ${1\over 2}$  \\
\hline
\end{tabular}
\label{smpart}}
\end{table}
\vskip .5in

\subsection{Anomaly Cancellation}

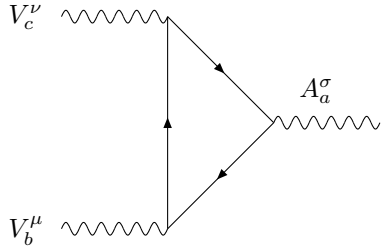
\begin{figure}[t]
\begin{center}
\begin{picture}(100,100)(-30,-5)
\SetScale{0.8}
\Photon(0,100)(50,100){3}{6}
\Photon(0,0)(50,0){3}{6}
\ArrowLine(50,100)(100,50)
\ArrowLine(100,50)(50,0)
\ArrowLine(50,0)(50,100)
\Photon(100,50)(150,50){3}{6}
\put(90,50){$A^\sigma_a$}
\put(-20,-3){$V^\mu_b$}
\put(-20,78){$V^\nu_c$}
\SetScale{1}
\end{picture}
\caption[]{Generic Feynman diagram contributing to gauge anomalies.  The
fermion loop contains all fermions transforming
 under the gauge symmetry. $a,b,c$ are gauge indices.
\label{fig:anomfig}}
\end{center}
\end{figure}

The requirement of gauge anomaly 
cancellation\cite{peskin,qu} 
 puts restrictions on the couplings of the fermions to
vector and axial gauge bosons, denoted here by $V^\mu$ and 
$A^\mu$.   
The fermions of the Standard Model 
have couplings to the gauge bosons of the general form:
\begin{equation}
{\cal L}\sim g_A {\overline \psi}
 T^\alpha \gamma_\mu\gamma_5\psi A^{\alpha\mu}
+g_V {\overline \psi} T^\alpha \gamma_\mu\psi V^{\alpha\mu},
\end{equation}
where $T^\alpha$ is the gauge generator in the adjunct representation.
These fermion-gauge boson couplings contribute to 
triangle graphs of the form shown in Fig. \ref{fig:anomfig}.  The 
triangle graphs
diverge at high energy,
\begin{equation}
T^{abc}
\sim Tr [\eta_i T^a \{ T^b, T^c\}] \int{d^nk\over (2\pi)^n}{1\over k^3},
\label{anomdef}
\end{equation}
where $\eta_i =\mp 1$ for left- and right-handed
fermions, $\psi_{L,R}={1\over 2} (1\mp \gamma_5)\psi$.
This divergence  is independent of the
fermion mass and depends only on the fermion couplings to
the gauge bosons.  Such
divergences cannot exist in a physical theory, and must somehow
be cancelled. 
The theory can be anomaly free in a vector-like model where the left-
and right-handed particles have identical couplings to gauge bosons and
the contribution to Eq. \ref{anomdef} cancels for each pair of
particles.
From Table 1, however, it is clear that the Standard Model is
not vector-like. The anomaly, $T^{abc}$, 
must therefore
be cancelled by a judicious choice
of fermion representations under the various gauge groups.

The only non-vanishing contribution to the
anomaly in the Standard Model is from
\begin{equation}
\Sigma Tr [Y \{T^a, T^b\}],
\label{anom_sm}
\end{equation}
where $T^a$ are the $SU(2)_L$ generators and the sum is over all
fermions in the theory.
Eq. \ref{anom_sm} vanishes for the hypercharge assignments given in Table 1.  
Note that the anomaly cancels separately for each generations of fermions.
The cancellation of the gauge anomalies in the Standard Model
for the third generation therefore
requires that the $b$ quark have a $T_3={1\over 2}$ partner with
electric charge $Q_{em}^t={2\over 3}\mid e\mid$, and 
hypercharge $Y^t=Q^t_{em}-T_3^t$.  The partner of the $b$ quark is by definition
the top quark.
  Since anomaly cancellation is independent
of mass, ${\it a~priori}$, the top quark mass could be anything.

\subsection{$b$ quark properties}
Many of the experimental properties of the $b$ quark require that it be a
$T_3^b=-{1\over 2}$ particle with $Q_{em}^b=-{1\over 3}$ 
and $Y^b={1\over 6}$.
  The coupling
of the $b$ quark  to the $Z$ boson can be tested to check if it has 
these quantum numbers.
The SM
fermions  couple to the $Z$ boson as,
\begin{equation}
{\cal L}=-{g\over 4 \cos \theta_W}{\overline \psi}
\gamma_\mu\biggl[ R_i (1+\gamma_5)+L_i(1-\gamma_5)\biggr] \psi_i Z^\mu~,
\end{equation}
where
\begin{eqnarray}
R_i&=& -2 Q_i \sin^2\theta_W\nonumber \\
L_i&=& 2 T_{3i} -2 Q_i \sin^2\theta_W~,
\label{zcoups}
\end{eqnarray}
and $\theta_W$ is the electroweak mixing angle.
The experimental value of $R_{had}$ is sensitive to $Q_{em}^b$,
\begin{equation}
R_{had}\equiv
{\sigma(e^+e^-\rightarrow {\hbox{hadrons}})\over
\sigma(e^+e^\rightarrow \mu^+\mu^-)}~.
\end{equation}
 At a center-of-mass energy, $\sqrt{s}=2 m_b$, $R_b$
depends sensitively on the $b$ quark electric 
charge,
\begin{equation}
\delta R_{had}(2m_b)=N_c (Q_b)^2 +{\cal O}(\alpha_s),
\end{equation}
where $N_c=3$ is the number of colors.
The experimental measurement,\cite{rbmeas}
\begin{equation}
\delta R_{had}(2m_b)=.36\pm.09\pm.03
\end{equation}
is in good agreement with the theoretical prediction
from  $Q^b_{em}=-{1\over 3}$, verifying the 
SM electric charge assignment of the $b$ quark.

Similarly, the $SU(2)_L$ quantum numbers of the
left- and right- handed  $b$ quarks
are probed by the decay rate for
$Z$ to $b {\overline b}$
 quark pairs.
If the $b$ did not have
a top quark partner, it would be
 an isospin $0$ particle ($T_3^b=0$) and the decay width
would be dramatically different from that of the SM.
The decay width is given in terms of the left and right-
handed couplings of the $Z$ to the $b {\overline b}$ pair,\cite{kuhn}
\begin{equation}
\Gamma(Z\rightarrow b {\overline b})={G_F M_Z^3
\over 4 \sqrt{2}\pi}(L_b^2+R_b^2).
\end{equation}
If the $b$ quark were an isospin singlet, then the decay 
width would be changed,
\begin{eqnarray}
{\Gamma(Z\rightarrow b {\overline b})^{T_3^b=-{1\over 2}}
\over
\Gamma(Z\rightarrow b {\overline b})^{T_3^b=0}}&=&
{1+4Q_b\sin^2\theta_W+8Q_b^2\sin^4\theta_W\over
8Q_b^2\sin^4\theta_W}\nonumber \\
&\sim & 13\quad .\nonumber
\end{eqnarray} 
The measurement of the $Z\rightarrow b {\overline b}$ decay width
excludes the $T_3^b=0$ hypothesis for the $b$ 
quark.\cite{ewtop}
\begin{itemize}
\item
The measured $b$ couplings, combined with anomaly cancelation, require
that the $b$ quark have a $T_3^t={1\over 2}$, color triplet, fermion
partner:  this is the top quark. 
\end{itemize}

\subsection{
Precision Measurements}

Before the top quark was discovered, an approximate value  for its
 mass was known from precision measurements,
 which depend sensitively
on the top quark mass.
At tree level, all electroweak measurements depend on just three parameters:
the $SU(2)_L\times U(1)_Y$ gauge coupling constants and the Higgs 
vacuum expectation value, $v=246~GeV$.
These are typically traded for the precisely measured quantities, $\alpha$,
$G_F$, and $M_Z$.  All electroweak measurements at lowest order
can be expressed in terms of these three parameters.
Beyond the lowest order, electroweak quantities depend on the masses
of the top quark and the Higgs boson.

A typical example of the role of the top
quark mass in precision measurements
is the calculation of the $\rho$ parameter,
\begin{eqnarray}
\rho & \equiv &  {M_W^2\over M_Z^2\cos^2\theta_W}
\nonumber \\
&=&{A_{WW}(0)\over M_W^2}-{A_{ZZ}(0)\over M_Z^2},
\label{rhodef}
\end{eqnarray}
where $A_{VV}$ is defined by the gauge boson $2$- point functions,
\begin{equation}
i\Pi_{VV}^{\mu\nu}\equiv A_{VV}(p^2)g^{\mu\nu}+B_{VV}p^\mu p^\nu
.
\end{equation}
At tree level, the $\rho $ parameter in the Standard Model
 is  exactly one, but at one loop it 
receives contributions from gauge boson, Higgs boson, and fermion loops.   The
largest corrections are those involving the top quark loop.
\begin{figure}[t]
\begin{center}
\hskip .75in
{{\epsfysize=1.in\epsffile
[ 52 293 542 408]{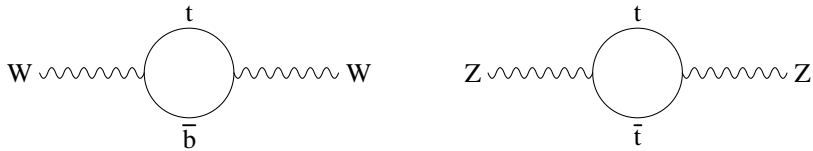}}}
\caption[]{Feynman diagrams for  gauge boson self-energies which give
contributions proportional to $M_t^2$.}
\label{fig:selfig}
\end{center}
\end{figure}
For simplicity, we compute only the
  corrections proportional to $M_t^2$, which  are found from the diagrams of
Fig. \ref{fig:selfig}.

\begin{figure}[t]
\begin{center}
\hskip .2in
{{\epsfysize=4.in\epsffile
[ 0 0 600  850]{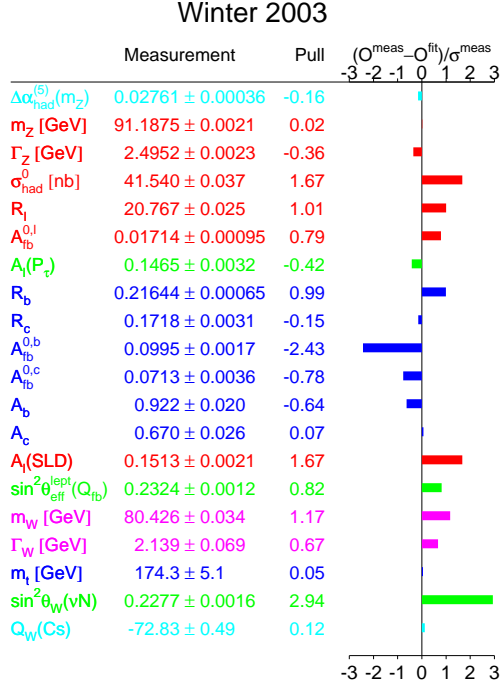}}} 
\caption[]{Fits to electroweak  data.  The bars show the
pull from the Standard Model global fit.\cite{ewtop}}
\label{fig:moriond}
\end{center}
\end{figure}

The 
2-point function of  the $Z$ boson is given by,
\begin{equation}
i\Pi_{ZZ}^{\mu\nu}=-N_c({-ig\over 4 
c_W})^2 (i)^2\int{d^nk\over (2 \pi)^n}
{T^{\mu\nu}\over [den]}
\end{equation}
 where $c_W\equiv \cos\theta_W$ and ,
\begin{eqnarray}
T^{\mu\nu}&=&Tr[(k+M_t)\gamma^\mu(R_tP_++L_t P_-)
 (k-p+M_t)
\gamma^\nu(R_tP_++L_tP_-)]
\nonumber \\
 den &=&[k^2-M_t^2][(k+p)^2-M_t^2]~,
\end{eqnarray}
and $P_\pm={1\over 2}(1\pm \gamma_5)$.
The coupling of the $Z$ to the top quark is given in Eq. 5 and 
$p$ is the external gauge boson momentum.
Shifting  momentum in the integral,
$k^\prime\rightarrow k-px$,  and using Feynman
parameters, the integral in $n=4-2\epsilon$ dimensions  becomes,
\begin{eqnarray}
i\Pi_{ZZ}^{\mu\nu}&=&
-N_c({-ig\over 4 c_W})^2 (i)^2\int{d^nk^{\prime}
\over (2 \pi)^n}
\int^1_0 dx
\nonumber \\
&& ~~~~
{[4k^{\prime~2}(-1+{\epsilon\over 2})(R_t^2+L_t^2)+16 L_tR_t
M_t^2]g^{\mu\nu}+ ...\over
(k^{\prime~2}+p^2 x(1-x)-M_t^2)^2}~,
\end{eqnarray}
where we retain only those terms contributing to $A_{ZZ}(0)$.
The result proportional
to $M_t^2$  is,
\begin{equation}
A_{ZZ}(p^2=0)={g^2N_c\over 32\pi^2c_W^2}\biggl(
{4\pi\over M_t^2}\biggr)^\epsilon
{M_t^2\over \epsilon}(R_t-L_t)^2.
\label{azz}
\end{equation}
The analogous result for the $W$ two-point function
 is,
\begin{equation}
A_{WW}(p^2=0)={g^2N_c\over 32\pi^2}
\biggl({4\pi\over M_t^2}\biggr)^\epsilon
M_t^2\biggl({1\over\epsilon}+{1\over 2}\biggr)~.
\label{aww}
\end{equation}
Combining Eqs. \ref{azz} and \ref{aww}, the contribution
to the $\rho $ parameter which is proportional to $M_t^2$ is found
from Eq. \ref{rhodef},\cite{sirlin}
\begin{equation}
\delta
\rho={g^2N_c\over 64\pi^2}{M_t^2\over M_W^2}={G_FN_c M_t^2\over 8
\sqrt{2}\pi^2}~.
\label{rhoans}
\end{equation}
Experimentally\cite{pdg}  $\rho=1.00126^{+.0023}_{-.0014}$\cite{pdg}  and
so an upper limit on the top quark mass can be obtained from 
Eq. \ref{rhoans}.

Many of the 
precision measurements  shown in Fig.
\ref{fig:moriond} are sensitive to $M_t^2$ and so a prediction
for the top quark mass can be extracted quite 
precisely by combining many measurements.
In fact, precision measurements were sensitive to
the  top quark mass
{\it before}
 top was discovered at Fermilab!
\begin{figure}[t,b]
\begin{center}
\hskip .25in
{{\epsfysize=3.in\epsffile
[ 0 0 624  510]{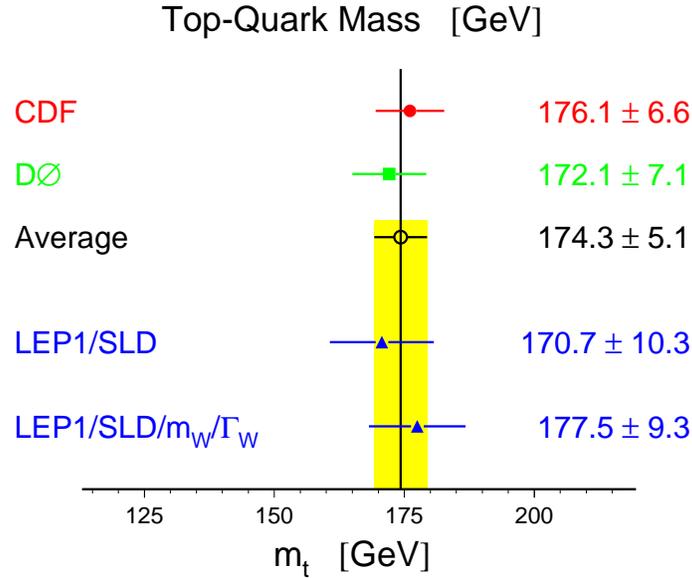}}} 
\caption[]{Measurements of the top quark mass at
Fermilab (CDF and D0) and indirect predictions
from precision measurements (LEP1, SLD and $M_W$).\cite{ewtop}}
\label{fig:topmass}
\end{center}
\end{figure}
The agreement between the direct measurement of the top
quark mass in the Fermilab collider experiments and
the indirect prediction from the precision measurements
(as shown in Fig. \ref{fig:topmass}) is one of the triumphs
of the Standard Model.  It is interesting to note that a 
similar limit on the Higgs boson mass from precision 
measurements gives $M_h< 193~GeV$ at  the $95\%$ confidence
level.\cite{ewtop}  Precision measurements depend logarithmically
on the Higgs mass, and so it is much more difficult to
bound the Higgs mass in this manner than it is to 
restrict the top quark mass.  Increases in the experimental precisions
of the top quark mass and the $W$ boson mass in Run II measurements
at the Tevatron will provide an improved bound on the Higgs boson mass,
$\delta M_h/M_h\sim 40\%.$\cite{higgsmass}

In the next sections, we discuss the discovery of the
top quark at Fermilab and the experimental exploration
of the top quark properties, both at Fermilab and the LHC and 
at a future
high energy $e^+e^-$ collider.

\section{Top Quark Properties}

\subsection{Hadronic Production}
The top quark was discovered in 1995 
at Fermilab in  $p {\overline p}$ collisions
 at $\sqrt{S}=1.8~TeV$.\cite{tdis}
This data set (called Run I)  consists of
an integrated luminosity of  
${\cal L}\sim 125~pb^{-1}$.  Both CDF and D0 are currently rediscovering
the top quark in the Run II data set.  Run II will produce roughly 500
clean top quark events for each inverse femtobarn of data and so precision measurements of many top quark  properties will be possible.

\begin{figure}[t]
\hskip 1in
{{\epsfysize=3.5in\epsffile
[ 0 0 624  510]{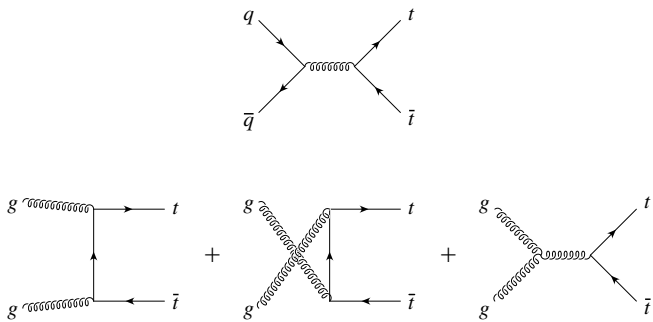}}} 
\vskip -1.5in.
\caption{Feynman diagrams contributing to top quark
pair production in hadron colliders.}
\label{fig:hadtt}
\end{figure}

In hadronic interactions, the top quark is  produced by
gluon fusion and by 
$ q {\overline q}$ annihilation as shown
in Fig. \ref{fig:hadtt},  
\begin{eqnarray}
gg &\rightarrow& t {\overline t}\nonumber \nonumber \\
q {\overline q} &\rightarrow& t {\overline t}.
\end{eqnarray}
%
%
The 
hadronic top quark production cross section, $\sigma_H$,  at the Tevatron, 
$p {\overline p}\rightarrow t 
{\overline t}$, (or $p p \rightarrow t
{\overline t}$ at the LHC) is found by convoluting
the parton level cross section with the parton
distribution functions (PDFs),
\begin{equation}
\sigma_H(S)=\Sigma_{ij}\int
f_i(x_1,\mu)f_j(x_2,\mu){\hat \sigma}_{ij}(x_1x_2S,\mu),
\label{eq:sigdef}
\end{equation}
where the
hadronic center of mass energy is $S$, the
partonic  center of mass 
energy is $s=x_1x_2 S$, the
parton distribution functions are $f_i(x,\mu)$,
and the parton level cross section is 
${\hat \sigma}_{ij}(s,\mu) $.
The parameter $\mu$ is an unphysical renormalization/factorization
scale.
 To ${\cal O}(\alpha_s^2)$, the sum in
Eq. \ref{eq:sigdef} is over the $q {\overline q}$ and $gg$ initial
states shown in Fig. \ref{fig:hadtt}.

   It is useful to
parameterize the parton level
 cross section (to next-to-leading order in $\alpha_s$) as,
\begin{equation}
{\hat \sigma}_{ij}(s,\mu)={\alpha_s^2(\mu)\over M_t^2}
\biggl[h^0_{ij}(\rho)+h^1_{ij}
\alpha_s(\mu)+h^2_{ij}\alpha_s(\mu)\log\biggl({\mu^2\over
M_t^2}\biggr)\biggr]~.
\label{sighad}
\end{equation}
At lowest order, ${\cal O}(\alpha_s^2)$, 
the parton level cross sections are,\cite{nde}
\begin{eqnarray}
h^0_{q {\overline q}}&=&{\pi\beta\rho\over 27} (2+\rho)\nonumber \\
h^0_{gg}&=& {\pi\beta\rho\over 192}
\biggl[ {1\over \beta}(\rho^2+16\rho+16)
\log\biggl({1+\beta\over 1-\beta}\biggr)-28-31\rho\biggr]
\nonumber \\
h^0_{qg}&=& 0 ~,
\label{lotop}
\end{eqnarray}
where 
\begin{eqnarray}
\rho&=&{4 M_t^2\over s}\nonumber \\
\beta &=& \sqrt{1-\rho}~.
\end{eqnarray}
The threshold condition is $\rho=1$ and  so ${\hat \sigma}_{ij}$ vanishes
at threshold.

To any given order in $\alpha_s(\mu)$, the hadronic
cross section $\sigma_H$ must be  independent
of $\mu$,
\begin{equation}
{d\sigma_H\over d \log\mu^2}=0.
\end{equation}
Applying this restriction to Eq. \ref{sighad} yields,
\begin{eqnarray}
0&=&\int
\biggl\{
 {\partial f(x_1,\mu)\over \partial \log\mu^2} f(x_2,\mu)
+f(x_1,\mu){\partial f(x_2,\mu)\over \partial \log\mu^2}
\biggr\}
{\hat \sigma}(x_1x_2S,\mu) dx_1 dx_2\nonumber \\
&& +\int f(x_1,\mu) f(x_2,\mu) 
{\partial {\hat\sigma}(x_1x_2S,\mu)\over \partial
\log\mu^2} dx_1 dx_2~.
\end{eqnarray}
The scale dependence of the PDFs is governed by the 
Altarelli-Parisi evolution functions, $P_{ij}(z)$:
\begin{equation}
{\partial f_i(x,\mu)\over \partial\log\mu^2}=
{\alpha_s\over 2 \pi}\Sigma_{j}
\int_x^1{dy\over y} P_{ij}\biggl({x\over y}\biggr)f_j(y,\mu)~.
\end{equation}
Similarly, the scale dependence of the strong 
coupling constant, $\alpha_s(\mu)$, is governed by the evolution
of the QCD $\beta$ function.
\begin{equation}
{\partial \alpha_s(\mu)\over \partial \log \mu^2}= -b_0 \alpha_s^2
\quad ,
\end{equation}
where $b_0= {(33-2n_{lf})\over 12 \pi}$, and $n_{lf}$ is
the number of light flavors.
At next-to-leading order the scale dependence of the 
various terms must cancel, yielding a prediction for
$h_{ij}^2$,
\begin{eqnarray}
h_{ij}^2&=& {1\over 2 \pi}\biggl[
4 \pi b_0 h_{ij}^0(\rho)-\Sigma_k
\int_\rho^1 dz h^0_{kj}\biggl({\rho\over z}\biggr)P_{ki}(z)
\nonumber \\
&&-\Sigma_k
\int_\rho^1 dz h^0_{ik}\biggl({\rho\over z}\biggr)P_{kj}(z)\biggr].
\nonumber \\
\label{scale}
\end{eqnarray}
Using the explicit forms of the Altarelli-Parisi evolution
functions, combined with Eq. \ref{lotop}, we find\cite{nde}
\begin{eqnarray}
h_{q {\overline q}}^2&=& {1\over 2\pi}\biggl[
{16\pi\rho\over 81}\log\biggl({1+\beta\over
1-\beta}\biggr) +{1\over 9} h_{q {\overline q}}^0(\rho)
\biggl(127
-6 n_{lf}+48\log\biggl({\rho\over 4 \beta^2}\biggr)
\biggr)\biggr]
\nonumber \\
h_{gg}^2&=&{1\over 2 \pi} \biggl[ {\pi\over 192} 
\biggl\{ 2\rho \biggl(59\rho^2+198\rho-288\biggr)
\log\biggl({1+\beta\over 1-\beta}\biggr)\nonumber \\
&&+12 \rho\biggl(\rho^2+16\rho+16\biggr) g_2(\beta)
-6\rho\biggl(\rho^2-16\rho+32\biggr)g_1(\beta)
\nonumber \\
&&-{4\over 15}\beta\biggl(7449\rho^2-3328\rho+724\biggr)
\biggr\}
-12 h_{gg}^0(\rho)\log\biggl({\rho\over 4\beta^2}\biggr)
\biggr]
\nonumber \\
h_{qg}^2&=& {1\over 384}\biggl[{4\rho\over 9}
\biggl(14\rho^2+27\rho-136\biggr)
\log\biggl({1+\beta\over 1-\beta}\biggr)-{32\over 3}
\rho(2-\rho)g_1(\beta)\nonumber \\
&&-{8\beta\over 135}\biggl(1319\rho^2-3468\rho+724\biggr)
\biggr]
\end{eqnarray}
where we have defined,
\begin{eqnarray}
g_1(\beta)&=&\log^2\biggl({1+\beta\over 2}\biggr)
-\log^2\biggl({1-\beta\over 2}\biggr)
+2 Li_2\biggl({1+\beta\over 2}\biggr)
-2Li_2\biggl({1-\beta\over 2}\biggr)\nonumber \\
g_2(\beta)&=& Li_2\biggl({2\beta\over 1+\beta}\biggr)
-Li_2\biggl({-2\beta\over 1-\beta}\biggr)~.
\end{eqnarray}

The quantities $h_{ij}^1$ can only be obtained by performing
a complete next-to-leading order 
calculation\cite{nde,ttqcd} and analytic
results are not available, although a numerical parameterization
is quite accurate.
The strong coupling evaluated at the top quark mass is small,
$\alpha_s(M_t) \sim 0.1$ , and so a
perturbative expansion converges rapidly.
At energy scales significantly different from the top quark mass, there
are also large logarithms of the form
 $(\alpha_s\log({M_t^2\over Q^2}))^n$, which can be summed to all
orders to obtain an improved prediction for the cross section.\cite{tt_nll}  The inclusion of next-to-leading-logarithm
effects reduces the scale dependence of the cross section
to roughly $\pm 5\%$.\cite{tnew}

\begin{figure}[t,b]
\begin{center}
\hskip .15in
{{\epsfysize=2.5in\epsffile
[-120  161 538  610]{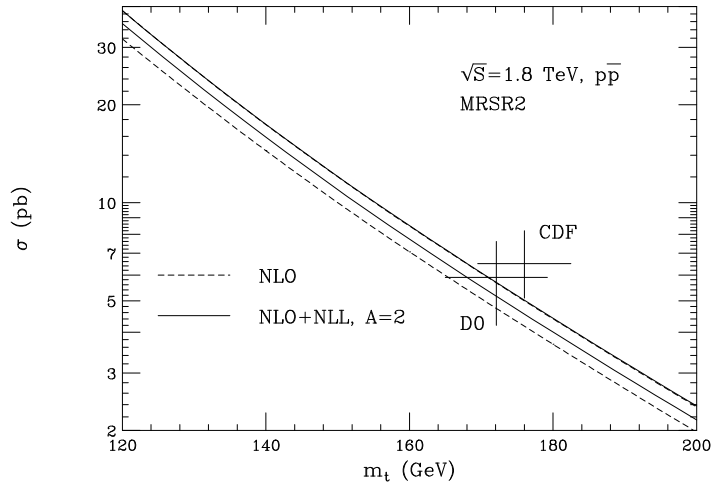}}} 
\caption[]{Fermilab Tevatron Run I cross sections for $p {\overline p}
\rightarrow t {\overline t}$. The dashed line is the NLO prediction, while
the solid lines incorporate the next-to-leading logarithms.\cite{tt_nll}}
\label{fig:tsig_fig}
\end{center}
\end{figure}

Fig. \ref{fig:tsig_fig} shows the Run I CDF and D0 measured top production cross
sections and top quark mass compared with two theoretical predictions.
The Run I $t {\overline t}$ cross sections are,\cite{tsig_ref}
\begin{eqnarray}
\sigma_{t {\overline t}}=&6.5^{+1.7}_{-1.4}~pb ~~~~&CDF  
\nonumber \\
\sigma_{t {\overline t}}=&5.9\pm {1.7}~pb ~~~~&D0~.
\end{eqnarray} 
The curves labelled NLO+NLL include some of the logarithms of the form
$(\alpha_s\log({M_t^2\over Q^2}))^n$ and the agreement with the experimental
results is clearly improved from the NLO prediction 
alone.  An updated theoretical study including 
next-to-leading-logarithm resummation gives the
prediction
\begin{equation}
\sigma_{t {\overline t}}(\sqrt{S}=1.8~TeV)=
(4.81-5.29)~ pb~ {\hbox{for}}~M_t=175~GeV,
\end{equation} 
with the range in  the prediction corresponding to
$M_t/2<\mu<2M_t$.\cite{tnew}
 The agreement between the predicted and the experimental
production rates  implies
that the top quark is a color triplet, since that rate would be significantly
different for a different color representation.

At the LHC,  the subprocess $gg\rightarrow t {\overline t}$ dominates the production rate with
$90\%$ of the total rate
and the
cross section is large, $\sigma_{LHC}\sim 800~pb$, yielding about
$10^8$ $t {\overline t}$ pairs/year.
Because of the large rate,
the LHC can search for new physics in top interactions.
The large rate also has the implication that
top production is often largest background to other
new physics signals.  A detailed understanding of the
top quark signal is therefore a necessary ingredient of new
physics searches.

\subsection{Weak Interactions of Top}

The 
weak interaction eigenstates are not mass eigenstates, 

\begin{equation}
{\cal L}=-{g\over 2\sqrt{2}}\Sigma_{q=d,s,b}
{\overline t} \gamma^\mu (1-\gamma_5)V_{tq}qW_\mu^+
+{\hbox{h.c.}}
\end{equation}
and  so the interaction of the top quark with a $W$ boson and a light
quark is proportional to the CKM mixing element, $V_{tq}$.
Measurements of the decay rates of the top quark into the
lighter quarks therefore translate directly into measurements of the
CKM mixing angles.

The 
unitarity of the CKM matrix  and the restriction to 
3 generations of fermions gives a limit on $V_{tb}$
from the measured values of $\mid V_{ub}\mid$ and $\mid
V_{cb}\mid$:\cite{pdg}
\begin{equation}
1= \mid V_{ub}\mid^2
+\mid V_{cb}\mid^2
+\mid V_{tb}\mid^2
\label{eq:unit_v}
\end{equation}
giving,
\begin{equation}
0.9991 <  \mid V_{tb}\mid < 0.9994~.
\label{eq:pdg_unit}
\end{equation}
If there are
more than 3 generations, however, Eq. \ref{eq:unit_v} is
no longer valid and there is almost no limit
from unitarity on $V_{tb}$.

Unitarity in the three generation model can
be tested by measuring the ratio of the rate for  top decays
to the $b$ quark to the rate for top decays to 
lighter quarks,\cite{cdf_unitarity}
\begin{eqnarray}
R_{tb }\equiv
{\Gamma(t\rightarrow Wb)\over
\Gamma(t\rightarrow Wq)}
&=&{\mid V_{tb}\mid^2
\over
\mid V_{td}\mid^2
+\mid V_{ts}\mid^2
+\mid V_{tb}\mid^2}
\nonumber \\
&& \nonumber \\
&=& .94^{+.31}_{-.24}~~~(CDF) 
\label{eq:cdf_unitmeas}
\end{eqnarray}
This quantity can be measured  at Fermilab by
counting the  number of tagged $b$'s in a 
top quark event.  Since the
$b$ quark 
lives approximately $1.5~ps$, the $b$ quark
will travel $450~\mu m$ before decaying.  The
secondary $b$ vertex can then be measured with 
a silicon vertex detector.  Assuming unitarity of the CKM matrix,
the denominator of Eq. \ref{eq:cdf_unitmeas} is $1$, and the measurement
can be interpreted as a measurement of $V_{tb}$,
\begin{equation}
\mid V_{tb}\mid =.97^{+.15}_{-.12},
\end{equation}
consistent with Eq. \ref{eq:pdg_unit}.

If there is a fourth generation of quarks,
$(t^\prime,b^\prime)$, the charge $-{1\over 3}$ $b^\prime$
is experimentally restricted to
 be heavier than $m_t-M_W$,\cite{pdg} and so 
the top quark cannot decay into the 
$b^\prime$.
Then the denominator of $R_{tb}$ need not equal $1$, and the measurement
implies only $\mid V_{tb}\mid >>\mid V_{ts}\mid ,~\mid V_{td}\mid$. 

The direct measurement of $V_{td}$  in single top production
will be discussed in Section 3.6.

\begin{figure}[t,b]
\begin{center}
\hskip 1in
{{\epsfysize=1.5in\epsffile
[236 300 366 403]{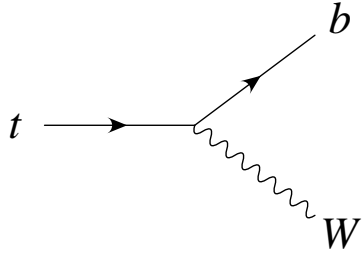}}} 
\caption[]{Feynman diagram for top decay to $W b$.}
\label{fig:tdecay_fig}
\end{center}
\end{figure}
\subsection{Top Quark Decay}
The top quark is the only quark which decays before it can form
a bound state.  This is due to the very short lifetime of the
top quark,
\begin{equation}
\tau_t\sim  5\times 10^{-25}~sec~,
\end{equation}
compared to the
QCD time scale:
\begin{equation}
 \tau_{QCD}\sim 3\times 10^{-24}~sec~.
\end{equation}
This is very different from the $b$ quark system, where the $b$
quark combines with the lighter quarks to form $B$ mesons, which
then decay.

Since $\mid V_{tb}\mid\sim 1$, the
 dominant decay  of the top quark is 
\begin{equation}
t(p)\rightarrow W^\mu(p_W)+ b(p^\prime),
\end{equation}
The lowest order amplitude is shown in Fig. \ref{fig:tdecay_fig} and is,
\begin{equation}
{\cal A}_0(t\rightarrow W^\mu b)
=-{g\over 2\sqrt{2}}{\overline u}(p^\prime)\gamma^\mu (1-\gamma_5)
u(p)~.
\label{eq:tdecay_rate}
\end{equation}
Squaring the amplitude, and summing over the $W$ polarization vectors,
\begin{equation}
\Sigma \epsilon^\mu(p_W)\epsilon^{*\nu}(p_W)
=-g^{\mu\nu}+{p_W^\mu p_W^\nu\over M_W^2},
\end{equation}
gives the amplitude-squared,
\begin{equation}
\mid {\cal A}_0(t\rightarrow Wb)\mid^2=
g^2(M_t^2-M_W^2)\biggl[1+{M_t^2\over 2 M_W^2}\biggr]~.
\end{equation}
The total decay width is 
\begin{equation}
\Gamma_0(t\rightarrow Wb)=\biggl({1\over 2} \biggr){1\over 2 M_t}
\int \mid {\cal A}_0(t\rightarrow Wb)\mid^2 (dPS_2) ~.
\end{equation}
The factor of ${1\over 2}$ is the average over the initial
top quark spin, and the phase space factor is,
\begin{equation}
(dPS_2)={M_t^2-M_W^2\over 8 \pi M_t^2} ~.
\end{equation}
Including the CKM mixing,
the final result is then,
\begin{equation}
\Gamma_0(t\rightarrow b W^+)\sim\mid V_{tb}\mid^2 {G_F M_t^3\over
8 \pi \sqrt{2}}\biggl(1-{M_W^2\over M_t^2}\biggr)^2
\biggl(1+{2M_W^2\over M_t^2}\biggr).
\end{equation}
Higher order QCD corrections can be calculated in a straightforward
manner and yield 
a precise prediction for the decay width,\cite{kuhn,dec}
 $$\Gamma(t\rightarrow b W^+)=
\mid V_{tb}\mid^2 1.42~GeV.
$$

\subsection{ $W$ Helicity in Top Quark Decay}
The 
helicity structure of the top quark decays is
interesting because it yields information on the $V-A$
nature of the $t b W$ vertex as is clear from Eq. 
\ref{eq:tdecay_rate}.
The top quark is produced through the
weak interactions as a left-handed fermion (neglecting
the $b$ quark mass), which 
has spin  opposite from  the direction of the
top quark motion.

If the $W$ is produced as a  longitudinal $W$ (helicity $0$), 
then its momentum and polarization vectors are
(see Fig. 8):
\begin{figure}[t]
\hskip 1in
{{\epsfysize=.5in\epsffile
[ 11 372 602 462]{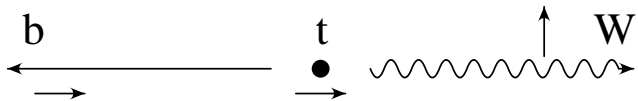}}} 
\label{longw}
\caption[]{Spin vectors for longitudinally polarized $W$'s produced
in the decay $t\rightarrow W b$.}
\end{figure}
\begin{eqnarray}
p_W&=& (E_W,0,0,\mid {\vec{p}_W}\mid)
\nonumber \\
\epsilon_L^W&=&{1\over M_W}(\mid {\vec{p}_W}\mid,0,0,E_W)
\nonumber \\
&\sim & {\mid {\vec{p}_W}\mid
\over M_W}+{\cal O}\biggl({M_W\over M_t}\biggr)~.
\end{eqnarray}
The amplitude for a top decaying into a longitudinal
$W$ is then,
\begin{eqnarray}
{\cal A}(t\rightarrow b W_L)&=&
-{g\over 2\sqrt{2}}
 {\overline b}{p_W\over M_W}(1-\gamma_5)t +{\cal O}\biggl({M_W\over M_t}\biggr)
 \nonumber \\
&=&-{m_t\over \sqrt{2}v}{\overline b}(1+\gamma_5)t
+{\cal O}\biggl({M_W\over M_t}\biggr) ~.
\end{eqnarray}
This gives the result that the decay of the top
into longitudinal $W$'s is,
\begin{equation}
\Gamma(t\rightarrow b W_L)\sim {m_t^3\over v^2}~.
\label{long_t_decay}
\end{equation}

The decay of the top quark into a positive helicity
$W$,
$h^W=+$, is forbidden by angular momentum conservation
since a massless $b$ quark  is always left-handed,
while the
heavy top can be 
either left- or right-handed.  This is illustrated
in Fig. 9.  CDF has measured the decay of the top quark to a 
right handed $W$ and finds a result consistent with $0$,\cite{tright}
\begin{equation}
BR(t\rightarrow b W_R)= .11\pm 0.15~.
\end{equation}

\begin{figure}[t,b]
\hskip 1in
{{\epsfysize=.5in\epsffile
[ 11 372 602 462]{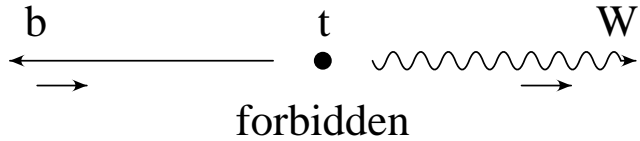}}} 
\label{tpol2}
\vskip .25in
\caption[]{Spin vectors for positive helicity
(right handed)  $W$'s produced
in the decay $t\rightarrow W b$. Conservation of angular momentum
forbids this decay.}
\end{figure}

\begin{figure}[h]
\hskip 1in
{{\epsfysize=.5in\epsffile
[ 11 372 602 462]{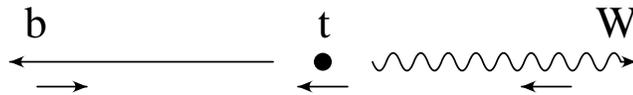}}} 
\label{fig:tpol2}
\caption[]{Spin vectors for transverse helicity  $W$'s produced in the
 decay $t\rightarrow W b$.}
\end{figure}
For a transversely polarized
$W$, the polarization vector is given by,
\begin{equation} 
\epsilon_{\pm}^W={1\over \sqrt{2}}(0,1,\pm i,0)~.
\end{equation}
This gives a decay rate,
\begin{equation}
\Gamma(t\rightarrow b W_T)\sim g^2 m_t~.
\label{trans_t_decay}
\end{equation}
Comparing Eqs. \ref{long_t_decay} and \ref{trans_t_decay}, it is
clear that
the rate for top decay to transverse $W$'s is 
suppressed relative to that for decay to longitudinal 
$W$'s,
\begin{eqnarray}
{\cal F}_0&=&{
\Gamma(t\rightarrow b W_L)\over \Gamma(t\rightarrow W b)_
{TOT}}
\nonumber \\&=& {{m_t^2\over 2 M_W^2}\over 1+{m_t^2\over 2 M_W^2}}
\nonumber\\
&=& 70.1\% ~.
\label{w0_pred}
\end{eqnarray}
Can we measure this?
The 
$W$ helicity is correlated with the momentum of the
decay leptons:
$h^W=+$  gives harder charged leptons than $h^W=-$.
A preliminary measurement of the polarization of
the $W$'s in top decay yields,\cite{tright}  
\begin{equation}
{\cal F}_0=91\pm 37 \pm 13\%~~CDF, 
\end{equation}
consistent with Eq. \ref{w0_pred}.
With
$2~fb^{-1}$ of data from
Run II at the Tevatron,  a more definitive ($\sim 5\%$) measurement
of ${\cal F}_0$ will be possible.\cite{run2_study}

\subsection{Top Production and Decay}

  The Tevatron and
the LHC produce predominantly
$t {\overline t}$ pairs, which decay almost entirely
by $t\rightarrow Wb$ (since $\mid V_{tb}\mid \sim 1$),\footnote{Recent
reviews of the experimental issues involved in top quark physics
can be found in Ref. \refcite{worm}.}
\begin{equation}
p {\overline p} \rightarrow t {\overline t}
\rightarrow
W^+W^-b {\overline b}.
\end{equation}
It is useful to classify top quark events by the decay pattern of
the $W^+W^-$ pair:
\begin{itemize}
\item
{\bf{Di-lepton}}:  Roughly $5\%$ of the $t {\overline t}$ events fall into
	this category,
\begin{eqnarray}
W^+ &\rightarrow & l^+\nu,~~l=e,\mu
\nonumber \\
W^- &\rightarrow & l^-\nu~. 
\end{eqnarray}
Since
$Br(W\rightarrow e\nu)+BR(W\rightarrow \mu\nu)\sim 22\%$, the
 $125~pb^{-1}$ at Run I of the Tevatron produced $\sim 625$ $t
{\overline t}$ events which yielded
$
\sim .22\times .22\times 625\sim 30
$ di-lepton events.  These events are 
clean, but suffer from the small statistics.  
\item
{\bf{
 lepton + jets}}:
This class of events comprises roughly $30\%$ of the total,
\begin{equation}
W^\pm \rightarrow l^\pm \nu,~~W^\mp \rightarrow jets ~.
\end{equation} 
Lepton plus jet events are
fully reconstructable and have small backgrounds.
Assuming at least 1 $b$ quark is  tagged,
CDF observed 34 such events in Run I, with an expected background
of 8.\cite{lepjets}  In 
Run II, with
 $2~fb^{-1}$, there should be
  1000 lepton+jets events/experiment.  This increase from Run I results
from
higher luminosity, the  higher cross section at $\sqrt{S}=2~TeV$
compared to $\sqrt{S}=1.8~TeV$, and upgraded detectors with improved
capabilities for $b$-tagging. 

This channel produced the most precise measurement of the cross
section for top pair production.  The combined average
from CDF and D0  at Run I is,
\begin{equation}
\sigma(p {\overline p}\rightarrow t{\overline t})\mid_{\sqrt{S}=1.8~TeV}
=5.9\pm 1.7~pb
\end{equation}
\item
{\bf{
All jets}}: $44\%$ of the events fall into this category.  This class of
events was observed by both CDF/D0 at Run
I, with 
large backgrounds.
\end{itemize}

\subsection{Single Top Production}
\begin{figure}[t]
{{\epsfysize=1.in\epsffile
[49 293 559 416]{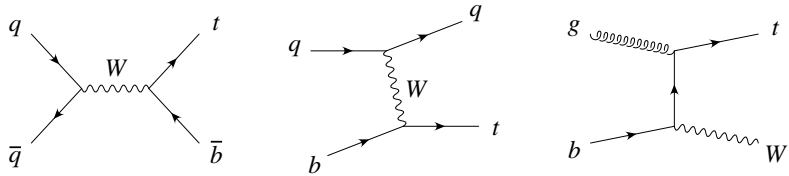}}}
\caption[]{Feynman diagrams contributing to single 
top production at the Tevatron
 and the LHC.}
\label{fig:single_top_fig}
\end{figure}

Single top production can provide a  precise measurement
of $V_{tb}$.  The dominant production processes at the Tevatron
are the $t$ channel $W$ exchange 
and 
 the $s$ channel $q {\overline q}$ annihilation
shown in Fig. \ref{fig:single_top_fig}.  At the LHC, the
subprocess $gb\rightarrow tW$ is also important.  The
contributions  (to NLO QCD) of the various subprocesses
at the Tevatron and the LHC are given
in Table 2.\cite{zack,stelzer}
  All of these processes are
proportional to $\mid V_{tb}\mid^2$.

\begin{table}[t]
\tbl{Contributions to single top production at the
 Tevatron and the LHC.}
{\footnotesize
\begin{tabular}
{|l|r|r|}\hline
 & Tevatron, RunII& LHC\\  \hline 
$\sigma_{qb}$ &  $2.1~pb $&  $238~pb$
\\ \hline  
$\sigma_{q{\overline q}}$ & $.9~pb$& $10~pb$\\ \hline  
 $\sigma_{gb}$ & $.1~pb$&  $56~pb$ \\ \hline
\end{tabular}
\label{singt}
}
\end{table}

Single top production was not observed in Run I, and 
both experiments set limits on the production rate.
At Run II of the Tevatron, and at the LHC, single top production
can be observed with a rate roughly half that
of the  $ t {\overline t} $
rate and will lead to a measurement at the Tevatron of,
$$
\mid{\delta V_{tb}\over V_{tb}}\mid
\sim 5\%
$$
with $30~fb^{-1}$ of data.  With $2~fb^{-1}$, the Tevatron Run II should
achieve an accuracy of
$\mid{\delta V_{tb}\over V_{tb}}\mid
\sim 13\%$.\cite{run2_study}

\subsection{
Measurements of $M_t$}
$M_t$ is  a fundamental parameter of the Standard Model,
but the value is not predicted (except indirectly from
precision measurements such as the $\rho$ parameter
discussed earlier).  In the 
Standard Model, a precise
value of $M_t$ is important primarily for its role in precision electroweak
observables and for the prediction for the Higgs boson mass extracted
from these measurements.\cite{higgsmt}  In extentions of the 
Standard Model, such as those discussed
in Section 5, a precise knowledge of
$M_t$  plays a critical role in defining
the parameters of the theory.

\begin{figure}[t]
\hskip 1in
\begin{center}
\hskip .2in
{{\epsfysize=3.5in\epsffile[0 0 600 600]
{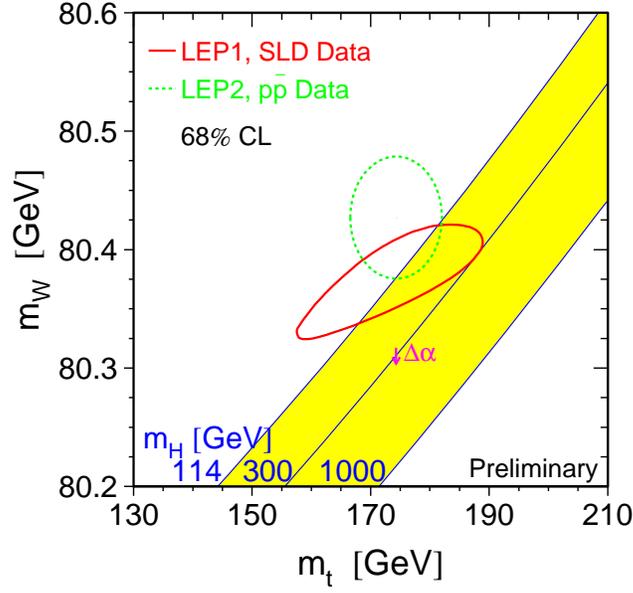}}}
\caption[]{Implications of precision measurements
of $M_t$ and $M_W$ on the indirect extraction of the
Higgs boson mass. The Standard Model allows only the shaded 
region.\cite{ewtop}}
\label{fig:mtmw_fig}
\end{center}
\end{figure}

The top quark mass was  measured in Run I at the Tevatron and the
most accurate value of $M_t$  comes from the single lepton+jets
channel,\cite{tmass}
\begin{eqnarray}
M_t &=&176.1\pm 4.8~ (stat)\pm 5.3~(syst)~~~CDF\nonumber \\
&=& 173.3\pm 5.6~(stat)\pm 5.5~(syst)~~~D0~~.
\end{eqnarray}
In the lepton plus jets channel there is one unknown
parameter (the longitudinal
momentum of the $\nu$ coming from the
$W$ decay) and three constraints from the reconstruction of
the $W$ masses and the requirement that  the reconstructed masses
of the top and anti-top be identical, $M_t=M_{\overline t}$.
The best value of the top quark mass found from combining all
channels at the Tevatron is,
\begin{equation}
M_t=174.3 \pm 5.1~GeV ~.
\end{equation}
The
systematic error is dominated by the  uncertainty on the
jet energy scale,
\begin{eqnarray}
\pm 4.4~GeV&&CDF\nonumber \\
\pm 4~GeV&&D0~~.
\end{eqnarray}
With increased data from Run II, the mass measurement
at the Tevatron will be improved to $\delta M_t
\sim 3~GeV$ with an integrated luminosity of $2~fb^{-1}$.

Fig. \ref{fig:mtmw_fig} shows the impact of a precise measurement of 
the top quark mass combined with a $W$ mass measurement for
the  prediction of the Higgs boson mass.  Clearly,
for the precision measurements of the $W$ and top masses to be
useful, the experimental error on both measurements
needs to be comparable. The Higgs mass prediction
extracted from precision measurements can be
compared with the direct measurement of the Higgs mass at the LHC
and the consistency of the Standard Model tested.   
Tevatron Run II with $2 fb^{-1}$ should achieve accuracies:
\begin{eqnarray}
 \delta M_W &\sim &27~MeV\nonumber \\ 
\delta M_t&\sim & 3~GeV~,
\end{eqnarray}
yielding a prediction for the Higgs mass with an uncertainty of\cite{higgsmt}
\begin{equation}
{\delta M_h\over M_h}\sim 40\% ~.
\end{equation}
The
LHC will have large statistics on top quark
production due to the large rate and can use the
lepton+jets channel with $10~fb^{-1}$ of data
  to obtain a measurement\cite{lhctop}
 \begin{equation}
\delta M_t\sim.07~GeV~~.
\end{equation}

\subsection{
Top Spin Correlations}

When pair produced in either $e^+e^-$ or $q
{\overline q}$ collisions, 
the  spins of  the $t{\overline t}$  pair are almost 
$100\%$ correlated.\cite{tspin}
The top decays before the spin can flip, and so spin
correlations between the $t$ and ${\overline t}$ result
in angular correlations among the decay products.  
At threshold, the $t {\overline t}$ cross section is 
 s-wave, and the $e^+e^-$ or 
$q {\overline q}$ spins are translated
directly to the $ t
{\overline t}$ spins.  The incoming
$q, {\overline q}$ have opposite helicities, since
they are interacting through a gauge 
interaction and the
$t {\overline t}$ spins aligned along the  beamline.
At
high energy, $\sqrt{s} >> m_t$, the  top mass 
is irrelevant and the
$t, {\overline t}$ must have opposite helicities.
Finding the  correct basis for intermediate energies is
subtle, however.  

\section{
The Top Quark  as a QCD Laboratory}

The production and decay mechanisms for the top quark 
in both $e^+e^-$ and hadronic collisions are well
understood and so the top quark system
 provides an ideal laboratory for studying
perturbative QCD.  This section contains a pedagogical treatment
of the ${\cal O}(
\alpha_s)$ next- to- leading order (NLO) QCD corrections  to the
process $e^+e^-\rightarrow t {\overline t}$.\cite{jlz}  
This  process is particularly simple since the QCD corrections
affect only the final state.
The calculation
is performed using the two cut-off phase space slicing (PSS)
algorithm for evaluating the real gluon
emission diagrams.\cite{pss,harris}
The total rate for  $e^+e^-\rightarrow
t {\overline t}$ tests our understanding of 
perturbative QCD and additionally
  the threshold behavior of the total cross section
can be used to obtain a
precise measurement of $M_t$, along with information
on $\alpha_s(M_t)$ 
and the $t {\overline t}h$ Yukawa coupling.  An understanding
of the QCD corrections are crucial for these interpretations.

As a second example of the role of QCD in top quark physics, we consider
the associated production of a $t {\overline t}$ pair along with a
Higgs boson, $h$.
The $t {\overline t}h$ Yukawa coupling is a fundamental
parameter of the theory and probes the mechanism of mass
generation, since in the Standard Model, $g_{tth}={M_t\over v}$.
  The processes $e^+e^-\rightarrow t {\overline t}h$
and $pp (p {\overline p})\rightarrow t {\overline t}h$ 
provide  sensitive measurements of the top quark- Higgs 
coupling and can be used to verify the correctness of the
Standard Model and  to search for new physics in the top quark and Higgs
sectors.
  The importance of the QCD NLO corrections to
$t {\overline t}h$ production
 is discussed at the end of this section.

\begin{figure}[t]
\begin{center}
\hskip -1in
\begin{picture}(100,100)(-30,-5)
\SetScale{0.8}
\ArrowLine(0,100)(50,50)
\ArrowLine(50,50)(0,0)
\Photon(50,50)(100,50){3}{6}
\ArrowLine(100,50)(150,100)
\ArrowLine(150,0)(100,50)
\put(140,90){$t(p^\prime)$}
\put(140, -5){${\overline t}(q^\prime)$}
\put(-50,90){$e^-(p)$}
\put(-50,-5){$e^+(q)$}
\put(55, 19){$\gamma,Z$}
\end{picture}
\caption{Lowest order
Feynman diagram for $e^+e^-\rightarrow t {\overline t}$.}
\label{fig:eetthfd}
\end{center}
\end{figure}
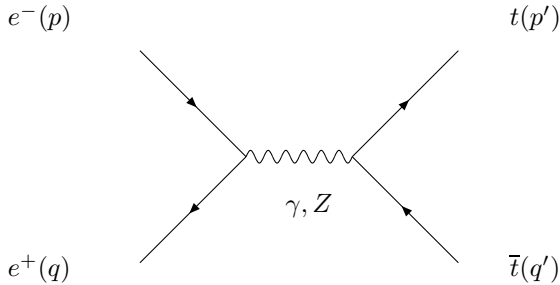

\subsection{NLO QCD corrections to $e^+e^-\rightarrow t
{\overline t}$}

The Born amplitude for
$e^+e^-\rightarrow t {\overline t}$ 
proceeds via $s-$ channel $\gamma$ and $Z$ exchange,
as shown in Fig. \ref{fig:eetthfd}.  The
dominant contribution at $\sqrt{s}=500~GeV$ is from $\gamma$
exchange and so for simplicity we will neglect the $Z$ contribution
here.  
(The complete ${\cal O}(\alpha_s)$ result, including the
contribution from $Z-$ exchange can be found in 
Refs. \refcite{jlz}
and \refcite{harris}.)  The electron mass can also be neglected.
The lowest order amplitude is:
$$
{\cal A}_{LO}(e^+e^-\rightarrow t {\overline t})
={e^2Q_eQ_t\over s} {\overline v}(q)
\gamma^\mu u(p){\overline u}(p^\prime)\gamma_\mu v(q^\prime)\delta_{ij}
~,
\label{eett_lo}
$$
where $i,j$ are the color indices of the outgoing top quarks.
It is simplest to work in the
center-of-mass frame of the incoming
$e^+e^-$ pair, where the $4-$ momenta of the particles
are given by:
\begin{eqnarray}  
p&=&{\sqrt{s}\over 2}(1,0,0,1)\nonumber \\
q&=&{\sqrt{s}\over 2}(1,0,0,-1)\nonumber \\
p^\prime&=&{\sqrt{s}\over 2}(1,0,\beta \sin\theta,\beta\cos\theta)\nonumber \\
q^\prime&=&{\sqrt{s}\over 2}(1,0,-\beta\sin\theta,-\beta\cos\theta)\nonumber
\end{eqnarray}
with
\begin{eqnarray}
s&=&(p+q)^2\nonumber \\
\beta&=&\sqrt{1-{4M_t^2\over s}}
\end{eqnarray}
 and $\theta$ is the angle of
the top quark emission with respect to the electron beam 
direction.   The threshold condition
is defined by
$\beta=0$.

It is straightforward to find the 
tree level amplitude-squared in $n=4-2\epsilon$ dimensions, (summed over colors),
\begin{equation}
\mid {\cal A}_{LO}\mid^2 = 64  \pi^2\alpha^2 Q_t^2 N_c
\biggl[2(1-\epsilon)+\beta^2(\cos^2\theta-1)\biggr]
~.\end{equation}
The color factor is
$N_c=3$.
The corresponding
total cross section is:
\begin{equation}
\sigma_{LO}=
{1\over 2 s}\int \mid {\cal A}_{LO}\mid^2 (dPS)_2.
\end{equation}
Integrating over the two body 
phase space, 
\begin{equation}
(dPS)_2={2^{2\epsilon}\over 16\pi}
\biggl({4\pi\over s}\biggr)^\epsilon
\beta^{1-2\epsilon}{1\over \Gamma(1-\epsilon)}\int_0^\pi
\sin^{1-2\epsilon}\theta d\theta,
\end{equation}
and including a factor of ${1\over 4}$ for the average over the
initial electron spins, gives a total rate,
\begin{eqnarray}
\sigma_{LO}(e^+e^-\rightarrow t {\overline t})
&=&
{\beta\over 32\pi s}{1\over 4} \int_{-1}^1
 d\cos\theta \mid {\cal A}_{LO}\mid^2
+{\cal O}(\epsilon)
\nonumber \\
&&\nonumber \\
&=& {2\pi N_c\alpha^2 Q_t^2\over 3 s} \beta (3-\beta^2)
+{\cal O}(\epsilon).
\nonumber
\end{eqnarray}
The ${\cal O}(\epsilon)$ terms which we have neglected will be
important in computing the ${\cal O}(\alpha_s)$
 QCD corrections.  
The cross section
vanishes at threshold, $\beta\rightarrow 0$, and
decreases with increasing energy, $\sigma \sim {1\over s}$.
This has the obvious implication that the higher energy an $e^+e^-$ collider,
the larger the required luminosity.

\subsection{
${\cal O}(\alpha_s)$ corrections to $e^+e^-\rightarrow
t {\overline t}$}

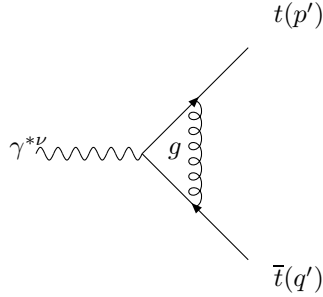
\begin{figure}[t,b]
\begin{center}
\hskip -1in
\begin{picture}(100,100)(-30,-5)
\SetScale{0.8}
\Photon(0,50)(50,50){3}{6}
\Gluon(75,75)(75,25){3}{6}
\ArrowLine(50,50)(100,100)
\ArrowLine(100,0)(50,50)
\put(-10,40){$\gamma^{*\nu}$}
\put(50,40){$g$}
\put(90,90){$t(p^\prime)$}
\put(90,-10){${\overline t}(q^\prime)$}
\end{picture}
\caption{Vertex correction to $e^+e^-\rightarrow t {\overline t}$.
The $\gamma^{*\nu}$ denotes an off-shell photon which is connected
to the initial $e^+e^-$ state.}
\label{fig:virtfig}
\end{center}
\end{figure}

The ${\cal O}(\alpha_s)$ corrections to $e^+e^-\rightarrow t 
{\overline t}$ contain both real and virtual contributions:
\begin{equation}
\sigma_{NLO}=\sigma_{LO}+\sigma_{virtual}+\sigma_{real}.
\end{equation}
The one-loop virtual contribution consists of a vertex correction
and a wavefunction renormalization of the top quark leg.
The vertex correction is shown in Fig. \ref{fig:virtfig}
  and can be parameterized 
as:
\begin{equation}
{\cal A}_{vertex}= e Q_e {\overline v}(q) \gamma_\nu u(p) {1\over s}
\Gamma^\nu(p^\prime, q^\prime)
\end{equation}
with,
\begin{equation}
\Gamma_\nu(p^\prime,q^\prime)=
(-eQ_t)\delta_{ij}g_s^2 C_F [{\cal N}]
{\overline u}(p^\prime)
\biggl[A{(p^\prime+q^\prime)_\nu \over
2 M_t}+B\gamma_\nu\biggr] v(q^\prime)
\label{eq:gamdef}
\end{equation}
where the normalization is,
\begin{equation}
[{\cal N}] \equiv {1\over 16 \pi^2}
\biggl({4\pi\mu^2\over M_t^2}\biggr)^\epsilon
\Gamma(1+\epsilon)
\end{equation}
and
\begin{equation}
C_F={N_c^2-1\over 2 N_c}={4\over 3} ~.
\end{equation}
 is the color factor for the vertex diagram.
The parameter $\mu$ is an arbitrary renormalization scale.
Explicit calculation gives the results,\cite{jlz,harris}
\begin{eqnarray}
A&=& (\beta-{1\over \beta})\Lambda_\beta
\nonumber \\
\Lambda_\beta &\equiv & \log\biggl({1-\beta\over 1+\beta}\biggr)\
\nonumber \\ 
&& \nonumber \\
B&=& {1\over \epsilon}
\biggl[1-({1\over\beta}+\beta)\Lambda_\beta\biggr]
-3\beta\Lambda_\beta
+\biggl({1\over\beta}+\beta\biggr)\biggl[-{1\over 2}
\Lambda_\beta^2+2\Lambda_\beta
\log\biggl({2\beta\over 1+\beta}\biggr)
\nonumber \\
&&+2Li_2({1-\beta\over 1+\beta})
+{2\pi^2\over 3}\biggr]
\nonumber\\
&\equiv &
{B_\epsilon\over \epsilon}+B_{fin}~~.
\nonumber
\end{eqnarray}
The di-logarithm is defined by,
\begin{equation}
Li_2(x)=-\int_0^1 dz {1\over 1-xz} ~.
\end{equation}

The ${\cal O}(\alpha_s)$ contribution to the
virtual cross section is  the interference
of the vertex correction with  the lowest order result.  (The $A$ term of Eq.~\ref{eq:gamdef} 
does not contribute to the interference with the
lowest order result when the electron mass is
neglected).
\begin{eqnarray}
\sigma_{vertex}&=&
2\biggl({1\over 2 s}\biggr)
\biggl({1\over 4}\biggr)
\int (dPS)_2 {\cal A}_{LO} {\cal A}_{vertex}^*
\nonumber \\
&=&{\alpha_s\over 2 \pi}
\biggl({4\pi\mu^2\over M_t^2}\biggr)^\epsilon C_F
\Gamma(1+\epsilon)\sigma_{LO}\biggl[
{B_\epsilon\over \epsilon}
+B_{fin}\biggr]~~.
\end{eqnarray}
Note that the ${\cal O}(\epsilon)$ terms from the lowest order
amplitude and from the 2-particle phase space combine with the
${B_\epsilon\over\epsilon}$ term to give a finite contribution
to $\sigma_{vertex}$.

\hskip 1.5in
\begin{figure}[t]
\begin{center} 
\hskip -1in
\begin{picture}(100,100)(-30,-5)
\SetScale{0.8}
\Photon(0,50)(50,50){3}{6}
\GlueArc(75,75)(15, 35, 240){3}{6}
\ArrowLine(50,50)(100,100)
\ArrowLine(100,0)(50,50)
\put(-10,40){$\gamma^{*\mu}$}
\put(90,90){$t(p^\prime)$}
\put(90,-10){$t(q^\prime)$}
\end{picture}
\caption{Feynman diagram for top quark wavefunction renormalization.}
\label{fig:ztdiag}
\end{center}
\end{figure}
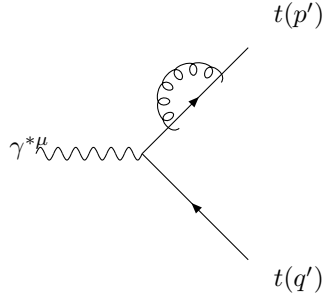
The
wavefunction renormalization for the top quark
 must also be included in  the
virtual correction, as shown in Fig. \ref{fig:ztdiag}.  This diagram gives
a multiplicative correction to the lowest order rate.
Using the on-shell renormalization scheme,
\begin{equation}
\sigma_Z=2\delta Z_{tt}\sigma_{LO},
\end{equation}
\noindent where
\begin{eqnarray}
\delta Z_{tt}&=&{\partial \Pi(p^2)\over \partial p^2}\mid p^2=M_t^2
\nonumber \\
&=&-{\alpha_s\over 4 \pi}
\biggl({4\pi\mu^2\over M_t^2}\biggr)^\epsilon
C_F\Gamma(1+\epsilon)\biggl({3\over \epsilon}+4\biggr)~.
\end{eqnarray}

The total virtual contribution is thus:\cite{jlz,harris}
\begin{eqnarray}
\sigma_{VIRT}&=&
\sigma_{vertex}+\sigma_Z
\nonumber \\
&=&
\sigma_{LO}{\alpha_s\over 2 \pi}
\Gamma(1+\epsilon) C_F
\biggl({4\pi\mu^2\over M_t^2}\biggr)^\epsilon
\biggl[(-3+B_\epsilon){1\over\epsilon}
-4+B_{fin}\biggr]~.
\label{eq:sigvirt}
\end{eqnarray}

\subsection{Real contributions}

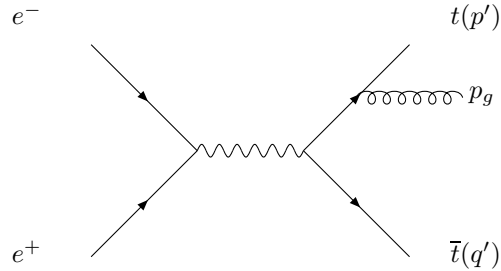
\begin{figure}[t]
\begin{center}
\hskip -1in
\begin{picture}(100,100)(-30,-5)
\SetScale{0.8}
\ArrowLine(0,100)(50,50)
\ArrowLine(0,0)(50,50)
\Photon(50,50)(100,50){3}{6}
\ArrowLine(100,50)(150,100)
\Gluon(125,75)(175,75){3}{6}
\ArrowLine(100,50)(150,0)
\put(-30,88){$e^-$}
\put(-30,-2){$e^+$}
\put(136,88){$t(p^\prime)$}
\put(143, 60){$p_g$}
\put(136,-2){${\overline t}(q^\prime)$}
\end{picture}
\caption{Sample Feynman diagram for real gluon emission
in the process $e^+e^-\rightarrow t {\overline t} g$.}
\label{fig:realfig}
\end{center}
\end{figure}

The 
virtual singularities
of Eq.~\ref{eq:sigvirt}  are cancelled by those arising from real gluon
 emission,
$e^+ e^-\rightarrow t {\overline t} g$.
A sample diagram of this type is shown in 
Fig. \ref{fig:realfig}.
The real gluon emission diagram of Fig. \ref{fig:realfig} contains a fermion
propagator on the outgoing top quark leg of the form:
$$
{1\over (p^\prime+p_g)^2-M_t^2}={1\over 2 p^\prime\cdot p_g}
={1\over 2 E^\prime E_g (1-{\hat \beta}\cos {\hat\theta})},
$$
where $E^\prime$ 
 and  $E_g$
 are the energies of the outgoing  top quark and gluon,
${\hat\theta}$ is the angle between them, 
and ${\hat \beta}=\sqrt{1-{M_t^2\over E^{\prime~2}}}$.
Because of the massive top quark, ${\hat \beta} \ne 1$,
 and this diagram has no
collinear singularity  as ${\hat\theta}\rightarrow 0$.
There is, however, a
soft singularity as the energy of the gluon vanishes,
 $E_g\rightarrow 0$.

The soft singularity is regulated 
using a technique called
phase space slicing (PSS).\cite{harris}  This method divides  
the gluon phase space into a soft gluon plus a hard gluon region:
\begin{equation}
\sigma_{real}\equiv\sigma_{soft}+\sigma_{hard}~.
\end{equation}
The soft region is defined by the energy of the emitted gluon,
\begin{equation}
E_g < \delta_s {\sqrt{s}\over 2},
\end{equation}
while the hard region is the remainder of the gluon phase space.
\begin{equation}
E_g >\delta_s {\sqrt{s}\over 2}~.
\end{equation}
The separation into hard and soft regions depends on the arbitrary
parameter $\delta_s$. Individually,
$\sigma_{soft}$ and $\sigma_{hard}$ depend on the cutoff, $\delta_s$,
and the
independence of the sum is  a  check on the calculation.  $\delta_s$
must be small enough that terms of ${\cal O}(\delta_s)$ can be
neglected, while large enough to prevent numerical instabilities.
 $\sigma_{hard}$ is 
finite and  can be evaluated numerically
using standard Monte Carlo techniques.

In the soft gluon regime, the cross section can be evaluated analytically.
The three body phase space is evaluated in  the  soft limit
in  the rest frame of the incoming $e^+e^-$:
\begin{equation}
p+q=p^\prime+q^\prime +p_g,
\end{equation}
giving a gluon energy of 
\begin{eqnarray}
E_g&=&\biggl({s-s_{p^\prime q^\prime}\over 2 \sqrt{s}}\biggr)\nonumber \\ 
s_{p^\prime q^\prime}&=&(p^\prime
+q^\prime)^2~.
\end{eqnarray}
The 3-body phase space in the soft limit can be
separated  into a 2 body phase space factor
 times a soft phase space factor:\cite{giele}
\begin{eqnarray}
(dPS)_3
&=&\int {d^{n-1}p^\prime\over 2 E_{p^\prime}
(2\pi)^{n-1}}
\int {d^{n-1}q^\prime\over 2 E_{q^\prime}
(2\pi)^{n-1}}
\cdot (2\pi)^n 
\delta^n(p+q-p^\prime-q^\prime -p_g)
\nonumber \\
&& ~~~~\cdot
\int {d^{n-1}p_g\over 2 E_g
(2\pi)^{n-1}}~~. 
\end{eqnarray}
For the  soft limit, we set the gluon momenta
$p_g=0$ in  the delta function and the phase space then factorizes:
\begin{equation}
(dPS)_3 (soft)
=(dPS)_2
\int {d^{n-1}p_g\over 2 E_g
(2\pi)^{n-1}} ~.
\end{equation}

The phase space is most easily evaluated by choosing an explicit
representation for the gluon momentum,
\begin{equation}
p_g=E_g(1,...,\sin\theta_1\sin\theta_2,
\sin\theta_1\cos\theta_2,\cos\theta_1).
\end{equation}
yielding,\cite{sterman}
\begin{eqnarray}
d^{n-1}p_g&=& d \mid {\vec p_g}\mid
\mid {\vec p_g}\mid^{n-2}d\Omega_{n-2}
\nonumber \\
&=& dE_g E_g^{n-2}
\sin^{n-3}\theta_1 d\theta_1 
\sin^{n-4}
\theta_2 d\theta_2 d\Omega_{n-4}
\end{eqnarray}
where,
\begin{equation}
\Omega_{n-4}=
{2\pi^{(n-3)/2}\over \Gamma({n-3\over 2})}.
\end{equation}
The integrals can be performed analytically to find the
three body phase space in soft limit:\cite{sterman}
\begin{eqnarray}
(dPS)_3(soft)&=& (dPS)_2
\biggl({4\pi\over s}\biggr)^\epsilon
\biggl({1\over 8 \pi^2}\biggr)
{\Gamma(1-\epsilon)\over \Gamma(1-2\epsilon)} dPS_{soft}\nonumber \\
&&\nonumber \\
dPS_{soft}&=&
{1\over \pi}\biggl({s\over 4}\biggr)^\epsilon
\int_0^{\delta_s{\sqrt{s}\over 2}}
dE_g E_g^{1-2\epsilon}
\nonumber \\ && ~~~~\cdot
\sin^{1-2\epsilon}\theta_1 d\theta_1 \sin^{-2\epsilon}
\theta_2 d\theta_2~.
\end{eqnarray}
Note that the soft phase space depends explicitely on the cut-off $\delta_s$
through the limit on the $E_g$ integral.

The soft phase space must be combined with the amplitude for
$e^+e^-\rightarrow t {\overline t} g$ in $p_g
\rightarrow 0$ limit: 
\begin{eqnarray}
{\cal A}_{soft}&=& {e^2 Q_e Q_t\over s} g_s T^A 
 {\overline v}(q)\gamma_\mu u(p)
\nonumber \\
&& ~~~~\cdot
{\overline u}(p^\prime)
\biggl\{\gamma^\sigma {(p^\prime+p_g+M_t)
\over (p^\prime+p_g)^2-M_t^2}\gamma^\mu
+\gamma^\mu
{(-q^\prime-p_g+M_t)
\over (q^\prime+p_g)^2-M_t^2}\gamma^\sigma\biggr\}
v(q^\prime)
\nonumber \\
&\rightarrow & g_s T^A
\biggl[{p^{\prime\sigma}\over p^\prime\cdot
p_g}-{q^{\prime\sigma}\over q^\prime\cdot
p_g}\biggr]{\cal A}_{LO}
\end{eqnarray}
where we have set $p_g=0$ in the numerator.  
The approximation of retaining only the most singular contributions is 
known as the
 eikonal ( or double pole)  approximation.\cite{eik_ref}
The soft limit  of the amplitude always factorizes in this manner
 into a factor multiplying the
lowest order amplitude.

The soft amplitude-squared is
\begin{eqnarray}
\mid {\cal A}_{soft}\mid^2 &=&g_s^2 Tr( T^A T^A) 
\mid {\cal A}_{LO}\mid^2  
\nonumber \\ &&
~~~~~\cdot\biggl[-{M_t^2\over (p^\prime\cdot p_g)^2}
 -{M_t^2\over (q^\prime\cdot p_g)^2}
+{s-2M_t^2\over p^\prime \cdot p_g
q^\prime\cdot p_g}\biggr]
\end{eqnarray}
giving the cross section in the soft limit,\footnote{As always, the
factor of $1/4$ is the spin average.}
\begin{equation}
\sigma_{soft}={1\over 2s}\biggl(
{1\over 4}\biggr) \int \mid {\cal A}_{soft}\mid^2 (dPS)_3(soft)~.
\end{equation}
The
problem reduces to evaluating the soft integrals.
A typical integral is 
\begin{equation}
I^{soft}\equiv
\int {1\over (p^\prime \cdot p_g)^2}dPS_3(soft)~.
\end{equation}
  Since the phase space and the amplitude squared
are Lorentz invariant, we are free to chose any
convenient frame.
We use
\begin{eqnarray}
p_g&=&E_g(1,...\sin\theta_1\sin\theta_2,
\sin\theta_1\cos\theta_2,\cos\theta_1)
\nonumber \\
p^\prime&=&{\sqrt{s}\over 2}(1,0,0,\beta)
\end{eqnarray}
and the soft integral is,
\begin{eqnarray}
I^{soft}&=&\int {1\over (p^\prime \cdot p_g)^2}dPS_3(soft)
\nonumber \\
&=& 
{1\over \pi}\biggl({s\over 4}\biggr)^\epsilon
\int_0^{\delta_s {\sqrt{s}\over 2}}
dE_g  E_g^{1-2\epsilon}
\sin^{1-2\epsilon}\theta_1 d\theta_1
\sin^{-2\epsilon}
d\theta_2\nonumber \\
&&\cdot 
\biggl({4\over s E_g^2}\biggr){1\over
(1-\beta \cos\theta_1)^2}
\nonumber \\
&& \nonumber \\
&=&
{2\over M_t^2}
\biggl(-{1\over 2\epsilon}+\log(\delta_s)
+{1\over 2 \beta}\Lambda_\beta\biggr)
+{\cal O}(\delta_s, \epsilon)~.
\end{eqnarray}
The complete set of soft integrals
relevant for this process is  given in the appendix of Ref. \refcite{rdw}.

The soft cross section is
\begin{eqnarray}
\sigma_{soft}&=&
\sigma_{LO}{\alpha_s\over 2 \pi} C_F
 \biggl({4\pi\mu^2\over s}\biggr)^\epsilon
{\Gamma(1-\epsilon)\over \Gamma(1-2\epsilon)}
\biggl[{C_{\epsilon}\over\epsilon}+C_{fin}\biggr],
 \end{eqnarray}
where
$B_\epsilon=-C_\epsilon$ so   the sum of virtual plus soft
contributions is finite.
The finite contribution,
$C_{fin}$, depends on $\delta_s$
and is given by\cite{harris}
\begin{eqnarray}
C_{fin}&=&-4\log(\delta_s)-{2\over\beta}\Lambda_\beta
-2\biggl({1\over\beta}+\beta\biggr)\nonumber \\&& \cdot
\biggl[Li_2\biggl({2\beta\over 1+\beta}\biggr)+{\Lambda_\beta^2
\over 4}+\log(\delta_s)\Lambda_\beta\biggr]
+{\cal O}(\delta_s)~.
\end{eqnarray}

The complete ${\cal O}(\alpha_s)$ result for $e^+e^-
\rightarrow t {\overline t}$ is then,
\begin{equation}
\sigma_{NLO}=
\sigma_{LO}+\sigma_{virtual}+
\sigma_{soft}+\sigma_{hard} ~.
\end{equation}
The combination $\sigma_{soft}+\sigma_{hard}$ is independent
of $\delta_s$, while the ${1\over \epsilon}$ poles cancel 
between $\sigma_{soft}+\sigma_{virtual}$.

This example, $e^+e^-\rightarrow t {\overline t}$, is simple enough 
that it could have been done analytically without the approximation
of phase space slicing\cite{jlz}.
Phase space slicing, however, is  useful for more complicated
examples and also 
to construct Monte Carlo distributions beyond the leading order.

\subsection{Threshold Scan in $e^+e^-\rightarrow t {\overline t}$}

\begin{figure}[t,]
\begin{center}
\hskip .75in
{{\epsfysize=2.5in\epsffile
[73 256 540 540]{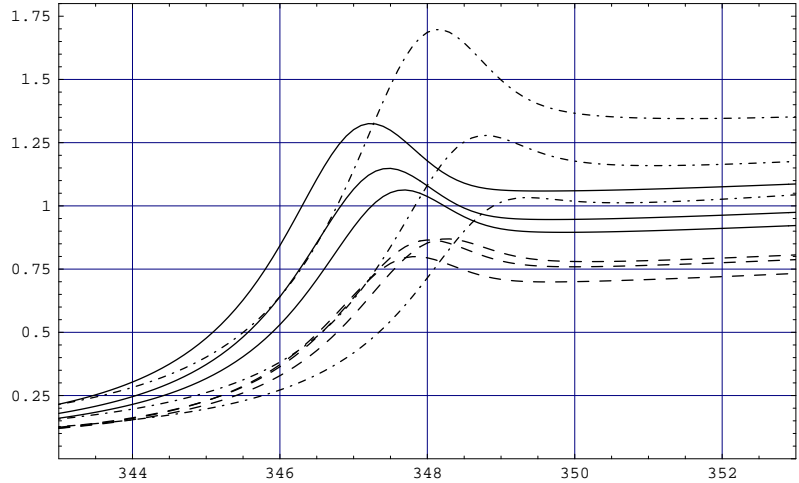}}} 
\caption[]{Cross section for $e^+e^-
\rightarrow t {\overline t}$ as a function of the
center-of-mass energy in GeV using the pole mass definition
for the top quark mass at lowest order ( dot-dashed), 
NLO (dashed), and NNLO  (solid). The vertical axis
shows $\sigma(e^+e^-\rightarrow t {\overline t})/\sigma_{pt}$,
where $\sigma_{pt}=4\pi\alpha^2/3s$.  
 The three sets of curves at
each order represent the variation with the renormalization
scale $\mu$, ($\mu=15,~ 30,$ and $60~GeV$).\cite{oleg}}
\label{fig:oleg1}
\end{center}
\end{figure}

The energy dependence of the cross section for 
$e^+e^-\rightarrow t {\overline t}$ at threshold depends sensitively
on the parameters of the top sector.
The location of the
peak depends on $M_t$,  while the height of the 
peak depends on $\Gamma_t$.  The 
problem is that the location of the peak as a function
of center-of-mass energy shifts at higher orders
and there is a  large renormalization
scale dependence to the prediction 
as seen in Fig.~\ref{fig:oleg1}.\cite{sumino}

At threshold, $\beta\rightarrow 0$, the 
theory possesses two small parameters: $\alpha_s$ and $\beta$
and the cross section diverges at ${\cal O}(\alpha_s)$,\cite{kuhn}
\begin{equation}
\sigma(e^+e^-\rightarrow
t {\overline t} )\mid_{\beta\rightarrow 0}=\sigma_{LO}
\biggl(1+{2\pi\alpha_s\over 3\beta}\biggr).
\end{equation}
This divergence  as $\beta \rightarrow 0$ is known as a
Coulomb singularity and  non-relativistic quantum mechanics
can be used to sum the leading contributions in powers of ${1\over
\beta}$.\cite{qn}

\begin{figure}[t]
\begin{center}
\hskip .75in
{{\epsfysize=2.5in\epsffile
[73 256 540 540]{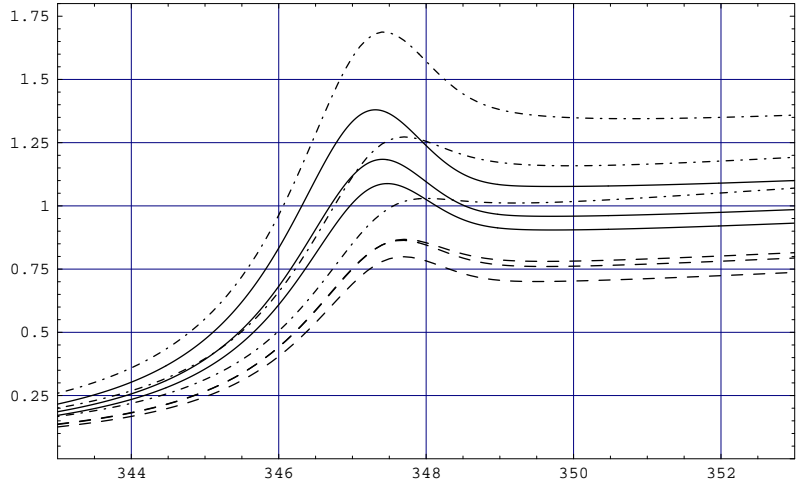}}} 
\caption[]
{Cross section for $e^+e^-
\rightarrow t {\overline t}$  as a function of the
center-of-mass energy in GeV using the 1S mass definition
for the top quark mass at lowest order ( dot-dashed), 
NLO (dashed), and NNLO  (solid). 
The vertical axis
shows $\sigma(e^+e^-\rightarrow t {\overline t})/\sigma_{pt}$,
where $\sigma_{pt}=4\pi\alpha^2/3s$.  
 The three sets of curves at
each order represent the variation with the renormalization
scale $\mu$, ($\mu=15,~ 30,$ and $60~GeV$). \cite{oleg}}
\label{fig:oleg2}
\end{center}
\end{figure}

Beyond lowest order in $\alpha_s$, 
the predictions are quite sensitive to the precise 
definition of the top quark mass.  The most straightforward definition is the
pole mass appearing in the top quark propagator, 
\begin{equation}
D(p)={i\over p-M_t\mid_{pole}-\Sigma(p)}~.
\end{equation}
The pole mass, $M_t\mid_{pole}$,
 is measured by reconstructing the four-momenta
of top quark decay products.
This definition of the top quark mass is, however, uncertain by QCD 
hadronization effects, ${\cal O}(\Lambda_{QCD})$. These
effects connect the top quark with its decay products.  Even more
disturbing is the fact that the location and height of the peak
in the $e^+e^-\rightarrow t {\overline t}$ cross section change
significantly between lowest order, NLO, and NNLO as
seen in Fig. \ref{fig:oleg1}.  A better definition of the
top quark mass is the
short distance $1S$ mass:\cite{oleg,onedef}
\begin{equation}
m_{1S}=M_t\mid_{pole}(1-{2\alpha_s^2\over 9}....)~.
\end{equation}
Using this mass definition, the cross section
for $e^+e^-\rightarrow t {\overline t}$
 peaks at $2m_{1S}$
and the location of the peak does not shift at NLO or NNLO, as
seen in Fig. \ref{fig:oleg2}.

The NNLO result still has roughly a $20\%$ scale uncertainty. 
 It is estimated
that by scanning the threshold dependence of the cross section 
for $e^+e^-\rightarrow t {\overline t}$ a
measurement with the 
accuracy
$\delta M_t\sim 200~MeV$ can be made with an integrated luminosity of
${\cal L}=50~fb^{-1}$.
This is to be compared with the expected
accuracy of the LHC measurement,
$\delta M_t\sim 1-2~GeV$.\cite{lhctop}  A precise measurement of the
top quark mass is of interest primarily because of its implications 
for electroweak precision measurements.\cite{higgsmt}

\subsection{$tth$ production in $e^+e^-$ and $pp$ collisions}

In many models with physics beyond the Standard Model, the Higgs coupling
to the top quark is significantly different from that in the Standard
Model. 
 In the Standard Model, the fermion-Higgs Yukawa couplings  are,
\begin{equation}
{\cal L}= - g_{ffh}{\overline \psi}\psi h
\end{equation}
where
\begin{equation}
g_{ffh}={M_f\over v}~.
\end{equation}
Since the top quark is the heaviest quark, it has the largest
Yukawa coupling to the Higgs boson.
It is important to measure this coupling accurately in order to
verify that the Higgs mechanism is the source of fermion masses.
 
The Higgs boson couplings to the lighter quarks can be 
tested by measuring the Higgs decays to fermion pairs.
This is not possible, however, for the top-fermion coupling.
Only for very heavy Higgs bosons ($M_h>2 M_t$) is the decay $h\rightarrow
 t {\overline t}$ kinematically accessible, and the branching ratio
of the Higgs into
$t {\overline t}$ pairs is significantly smaller than that into gauge
boson pairs.  It does not appear feasible to measure $g_{tth}$
through this channel.
The most promising prospect for  measuring
the  top quark Yukawa coupling
is   $t {\overline t} h$ 
production 
 in $e^+e^-$ and $pp$ (or $p {\overline p}$) collisions.

\subsection{
 $t {\overline t}h$ at an  $e^+e^-$ collider}

At an $e^+e^-$ collider, $t {\overline t} h$ associated production occurs
through $s$-channel $\gamma$ and $Z$ exchange, as shown in Fig. 19.
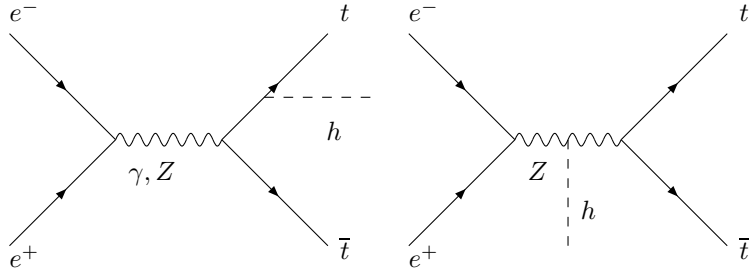
\begin{figure}[t]
\begin{picture}(100,100)(-20,-5)
\SetScale{0.8}
\ArrowLine(0,100)(50,50)
\ArrowLine(0,0)(50,50)
\Photon(50,50)(100,50){3}{6}
\ArrowLine(100,50)(150,100) 
\ArrowLine(100,50)(150,0)
\DashLine(120,70)(170,70){5} 
\put(45,25){$\gamma, Z$} 
\put(0,85){$e^-$}
\put(0,-8){$e^+$}
\put(120,40){$h$}
\put(125,85){$t$}
\put(125,-5){${\overline t}$}
\SetScale{1}
\end{picture}
\hskip -1in
\begin{picture}(100,100)(-140,-5)
\SetScale{0.8}
\ArrowLine(0,100)(50,50)
\ArrowLine(0,0)(50,50)
\Photon(50,50)(100,50){3}{6}
\DashLine(75,50)(75,0){5}
\ArrowLine(100,50)(150,100) 
\ArrowLine(100,50)(150,0)
\put(45,25){$Z$} 
\put(0,85){$e^-$}
\put(0,-8){$e^+$}
\put(65,10){$h$}
\put(125,85){$t$}
\put(125,-5){${\overline t}$}
\end{picture} 
\label{fig:eetth}
\caption[]{ Contributions to $e^+e^-\rightarrow t {\overline t}h$ at
lowest order.}
\end{figure}
The contribution 
from $Z$ exchange  is a few percent of the total rate and will be neglected 
here.  With this assumption, the rate is directly proportional to $g_{tth}^2$.
In order to interpret the cross section 
as a measurement of $g_{tth}$, however,  it
is necessary to include the QCD corrections
which are potentially of the same order of magnitude as new
physics effects which may change $g_{tth}$ from the SM prediction.
The NLO QCD result includes virtual and real gluon corrections on the
outgoing $t,{\overline t}$ legs.  The calculation follows that of the
previous section for $e^+e^-\rightarrow t {\overline t}$, although in 
this case the virtual calculation also includes box diagrams.
  These corrections have
been found in Refs. \refcite{eetth_QCD1}
and \refcite{eetth_QCD2} and can be parameterized as
\begin{equation}
K\equiv{\sigma_{NLO}\over \sigma_{LO}}=1+\alpha_s(\mu) F(s,M_t^2, M_h^2)~~.
\end{equation}
The parameter $\mu$ is an unphysical renormalization
scale and, at this order, the only dependence on $\mu$
is in the running of
$\alpha_s(\mu)$.  For $M_h=120~GeV$, and $\mu$ ranging from 
$M_t$ to $\sqrt{s}$,
\begin{eqnarray}
K(\sqrt{s}=500~GeV)&\sim& 1.4-1.5 \nonumber \\
K(\sqrt{s}=1~TeV)&\sim& .8-.9~.
\end{eqnarray}
The
NLO results for $t{\overline t}h$
production in $e^+e^-$ collisions are shown in Fig. \ref{fig:eenlo}.
The rate is quite small and has a 
maximum value for $\sqrt{s}\sim 700-800~GeV$ for $M_h$ in the $120~GeV$
region.
\begin{figure}[t]
\hskip 1in.
\begin{center}
{{\epsfysize=4.0in\epsffile
[ 0 0 600  600]{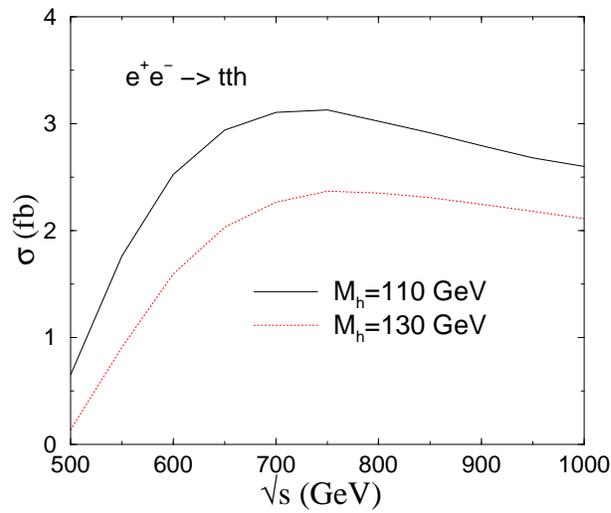}}} 
\caption[]{NLO rate for $e^+e^-\rightarrow t {\overline t} h$ as
a function of center of mass energy.\cite{eetth_QCD1}}
\label{fig:eenlo}
\end{center}
\end{figure}

In order to measure the
top quark Yukawa coupling in $e^+e^-$ collisions, we consider the final
state 
\begin{equation}
e^+e^-\rightarrow t {\overline t} h\rightarrow
W^+W^- b {\overline b} b {\overline b}~.
\end{equation}
The events are then classified according to the $W$ decays, in an
analogous manner to that discussed in Section 3.5
 for top quark decays.
Since the event rate is quite small, it is advantageous to
combine hadronic and semi-leptonic  $W$ decay channels.\cite{juste,baer}
At
$\sqrt{s}=800~GeV$ , an integrated luminosity 
of ${\cal L}=1000~fb^{-1}$, 
can measure the coupling to an accuracy of\cite{juste}
\begin{equation}
{\delta g_{tth}\over g_{tth}}\sim 5.5\%~.
\end{equation}
At a lower energy,
$\sqrt{s}=500~GeV$,  the rate is considerably smaller
and even with an integrated luminosity of
${\cal L}=1000~fb^{-1}$, 
the precision is quite poor,\cite{baer}
\begin{equation}
{\delta g_{tth}\over g_{tth}}\sim 20\%~,
\end{equation}
for 
$M_h=120~GeV$.

\subsection{$pp\rightarrow t {\overline t}h$ at NLO}

The
QCD corrections to
$p {p}  \rightarrow t {\overline t}h $
(or $p {\overline p}  \rightarrow t {\overline t}h $)
 can be calculated in
a similar fashion to those for $e^+e^-\rightarrow t 
{\overline t}h$.  In the hadronic case, however, the calculation is
complicated by the existence of strong interactions in the initial state.
In this section, we discuss the treatment of initial state singularities
in hadron collisions.
The
ingredients of an NLO calculation for $pp$ or $p {\overline p}
\rightarrow t {\overline t} h$ are:\cite{rdw,pptth_qcd}

\begin{itemize}
\item
$\sigma_{virtual}:$
This contains all of the one-loop diagrams contributing to the process.
Finite diagrams, or diagrams which contain only ultraviolet singularities,
can be evaluated numerically.\cite{vanold}
Diagrams which contain infrared singularities 
must be evaluated analytically.

\item
$\sigma_{soft}$:  
This contains the  soft singularities arising 
from both the 
initial and final state soft gluon radiation in the limit
$p_g\rightarrow 0$.
The calculation is analogous to the discussion of Section 4.3.

\item
$\sigma_{hard}:$
This is the contribution from real gluon radiation when the gluon
has an energy above the soft cut-off.  In contrast to the $e^+e^-$
example of Section 4.3, this contribution has a singularity when the initial state
quark or 
gluon is parallel to the radiated final state gluon.  The initial
state collinear singularities are absorbed into the 
parton distribution functions 
(PDFs).
\end{itemize}
The combination
$\sigma_{soft}+\sigma_{hard}+
\sigma_{virtual}$ is finite.

\begin{figure}[t]
\begin{center}
\hskip .25in
{\epsfysize=3.in
\epsffile[57 29  525 469]{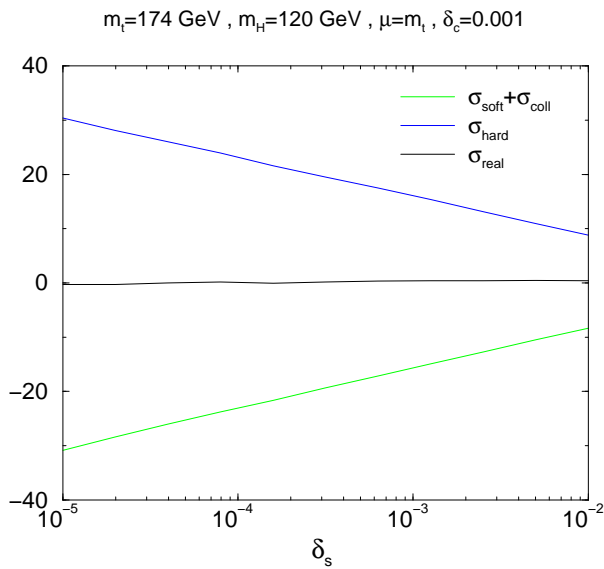}}
\caption[] {NLO QCD corrections to $p {\overline p}
\rightarrow t {\overline t} h$ at $\sqrt{S}=2~TeV$. This
figure shows the independence of the total cross section
on the soft cutoff, $\delta_s$.\cite{rdw}}
\label{fig:pptth_nlo}
\end{center}
\end{figure}

\subsection{Collinear Singularities and Phase Space Slicing}

To illustrate the effects of collinear singularities, we
consider the parton level subprocess,
\begin{equation}
q(p)+ {\overline q}(q)
 \rightarrow t(p_t)+ {\overline t}(p_{t^\prime})
+ h(p_h)+ g(p_g)~.
\end{equation}
As in the $e^+e^-$ example,
we  divide the gluon
phase space into  a soft and a hard region.
We then
further divide $\sigma_{hard}$
into a region with collinear divergences and a finite region,
\begin{equation}
\sigma_{hard}\equiv\sigma_{hard}\mid_{coll}+\sigma_{hard}\mid_{not-coll}
\end{equation}
The 
separation depends on a new cut-off, $\delta_c$, although
$\sigma_{hard}$ must be independent of $\delta_c$.  
If the emitted gluon is from the initial state quark line,
then there is a  quark propagator,
\begin{equation}
{1\over (p-p_g)^2}=-{1\over 2 p\cdot p_g}=-{1\over \sqrt{s}
E_g(1-\cos\theta)}
\label{singdef}
\end{equation}
which becomes singular when $p$ and $p_g$ are parallel
($\cos\theta\rightarrow 1$).
The
hard/not-collinear region is defined by,
\begin{equation}
{2 p\cdot p_g\over E_g \sqrt{s}}>\delta_c\qquad \qquad
{2 q\cdot p_g\over E_g \sqrt{s}}>\delta_c~.
\end{equation}
In this region the quark propagator
of Eq. \ref{singdef} is finite and
 the cross section can be computed numerically.

In the collinear region, 
\begin{equation}
{2 p\cdot p_g\over E_g \sqrt{s}}<\delta_c \qquad \qquad
{2 q\cdot p_g\over E_g \sqrt{s}}<\delta_c,
\end{equation}
and the
matrix element  factorizes
into a
parton $i$ splitting into a
 hard parton $i^\prime$ plus a collinear gluon with kinematics
\hskip .5in
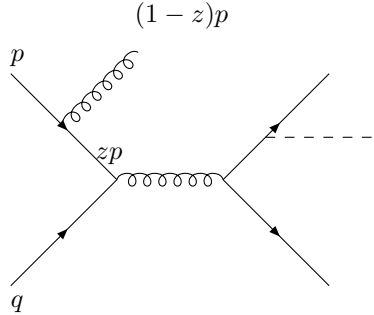
\begin{figure}[t]
\begin{center}
\begin{picture}(100,100)(-20,-5)
\hskip .5in
\SetScale{0.8}
\ArrowLine(0,100)(50,50)
\Gluon(25,75)(60,110){3}{6}
\ArrowLine(0,0)(50,50)
\Gluon(50,50)(100,50){3}{6}
\ArrowLine(100,50)(150,100) 
\ArrowLine(100,50)(150,0)
\DashLine(120,70)(170,70){5}  
\put(0,85){$p$}
\put(32,48){$zp$}
\put(47,100){$(1-z)p$}
\put(0,-8){$q$}
\SetScale{1}
\end{picture}
\label{fig:feyn_coll}
\caption[]{Diagram contributing to collinear singularity
for the process $q {\overline q}\rightarrow t {\overline t} h g$.}
\end{center}
\end{figure}
\begin{eqnarray}
p_{i^\prime} &=& z p_i\nonumber \\
p_g&=&(1-z) p_i
\nonumber\\
s_{ig}&\equiv & 2 p_i\cdot p_g\nonumber
\end{eqnarray}
These kinematics are illustrated in Fig. 22.
In the
current example, $q\rightarrow q g$
and the matrix element factorizes,\cite{rdw}
\begin{equation}
\mid
{\cal A}(ij\rightarrow t {\overline t} h+g)\mid^2
\rightarrow _{collinear}
\Sigma_{i^\prime}
g_s^2
\mid{\cal A}_{LO}(i^\prime j\rightarrow t {\overline t}h)
\mid^2
{2 P_{i i^\prime}(z)\over z s_{ig}}
\end{equation}
where $P_{i i^\prime}$ is the probability that
parton $i$ splits into parton $i^\prime$, with
fraction $z$ of the initial momentum.  In this example,
\begin{equation}
P_{i i^\prime}(z) = P_{qq}(z)=C_F\biggl[
{1+z^2\over 1-z}
-\epsilon (1-z)\biggr]~.
\label{colamp}
\end{equation}
The phase space also factorizes in the collinear limit:\cite{giele}
\begin{eqnarray}
&&(PS)_4(ij\rightarrow t {\overline t} h+g)_{collinear}
\rightarrow 
(PS)_3 (i^\prime j\rightarrow t {\overline t} h)
{\Gamma(1-\epsilon)\over \Gamma(1-2\epsilon)}
{(4\pi)^\epsilon\over 16 \pi^2}
\nonumber \\
&&~~~~~~~~~\cdot
 z~ dz~ ds_{ig}
[(1-z)s_{ig}]^{-\epsilon}
\theta\biggl({(1-z)\over z} p_{i^\prime}
\cdot p_j \delta_c-s_{ig}\biggr)
\label{colps}
\end{eqnarray}
Combining Eqs. \ref{colamp} and \ref{colps}, the 
hard-collinear parton level cross section becomes:\cite{colin}
\begin{eqnarray}
&&\sigma_{hard/coll}= {\alpha_s\over 2 \pi}
{\Gamma(1-\epsilon)\over \Gamma(1-2 \epsilon)}
\biggl({4\pi\mu^2\over M_t^2}\biggr)^\epsilon
\biggl(-{1\over \epsilon}\biggr)\delta_c^{-\epsilon}
\nonumber \\
&&~~~~~~~~
\cdot\Sigma_{i^\prime}
\biggl\{ \int_0^{1-\delta_s} dz
\biggl[{(1-z)^2 p_{i^\prime}\cdot p_j
\over  z M_t^2}\biggr]^{-\epsilon}
P_{i i^\prime}(z)\sigma_{LO}(i^\prime j\rightarrow
t {\overline t} h)
\nonumber \\
&&~~~~~~~~+(i\leftrightarrow j)\biggr\}~.
\label{eq:colres}
\end{eqnarray}
This result has a dependence on both 
$\delta_c$ and $\delta_s$.
The
$1-\delta_s$ limit on the  $z$ integration 
excludes soft gluon region.
The remaining singularities in Eq. \ref{eq:colres}
are absorbed into the mass factorization as described
in the next section.

\subsection{Mass Factorization}

The
 hadronic cross section depends on the PDFs, $f_i(x)$:
$$
\sigma_H(S)=\Sigma_{ij}\int
f_i(x_1)f_j(x_2){\hat \sigma}_{ij}(x_1x_2S) dx_1 dx_2~,
$$
where ${\hat \sigma}$ is the parton level cross section.
At lowest order, there is no dependence on the renormalization/
factorization scale, $\mu$.
Since the only physical quantity is the hadronic cross section,
$\sigma_H(S)$,  we can
define scale dependent PDFs to absorb the  hard/collinear singularity
of Eq. \ref{eq:colres},\cite{harris,giele}
\begin{eqnarray}
f_i(x,\mu)& \equiv&
f_i(x)+\biggl(-{1\over \epsilon}\biggr)
{\alpha_s\over 2 \pi}
{\Gamma(1-\epsilon)\over \Gamma(1-2 \epsilon)}
\biggl({4\pi\mu^2\over \mu^2}\biggr)^\epsilon
\nonumber \\
&& \cdot\biggl\{
\int_z^{1-\delta_s} {dz\over z}P_{i i^\prime}(z) f_{i^\prime} ({x\over z})
+C_F(2\log(\delta_s)+{3\over 2})\biggr\}~.
\label{eq:massfac} 
\end{eqnarray}
The upper limit on the $z$ integration excludes the
soft region.
After convoluting the parton cross section with the renormalized quark
distribution function of Eq.~(\ref{eq:massfac}), the infrared singular
counterterm of Eq.~(\ref{eq:massfac}) exactly cancels the remaining
infrared 
poles of $\hat{\sigma}_{virt}+\hat\sigma_{soft}+
\hat{\sigma}_{hard/coll}$.

The cancellation of the cutoff dependence at the level
of the total NLO cross section is a very delicate issue, since it
involves both analytical and numerical contributions. It is crucial to
study the behavior of $\sigma_{NLO}$ in a region where the cutoff(s) are
small enough to justify the approximations used in the analytical
calculation of the infra-red divergent soft and collinear
limits, but not
so small to give origin to numerical instabilities.  This
effect is illustrated in Fig. 21.

The NLO result for
$p {\overline p}\rightarrow
t {\overline t} h$ at  the Tevatron is shown in 
Fig. 23 as a function of the unphysical factorization/
renormalization scale.  The complete calculation includes not
only the $q {\overline q}$ initial state described here, but 
also the $gg$ and $qg$ initial states.
\begin{figure}[t]
\begin{center}
\hskip .25in
{\epsfysize=4.in
\epsffile[0 0 600 600]{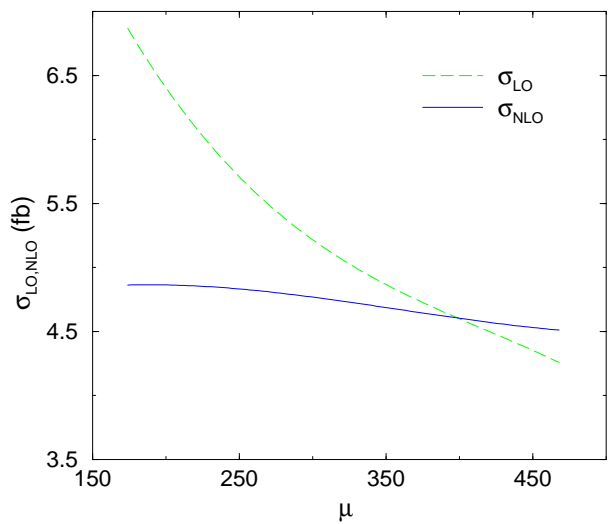}}
\label{fig:tevfig}
\caption[]{Leading order and next-to-leading order rate for
$p {\overline p}\rightarrow t {\overline t} h$ at the Tevatron for
$M_h=120~GeV$ and $\sqrt{S}=2~TeV$ as a function of the unphysical
factorization/renormalization scale.\cite{rdw}}
\end{center}
\end{figure}
Note the reduced $\mu$ dependence of the  NLO rate from the lowest
order prediction. The cross section is very small, making this 
a challenging measurement.

\begin{figure}[t]
\hskip .5in{\epsfysize=3.in\epsffile[6 15 278 279] {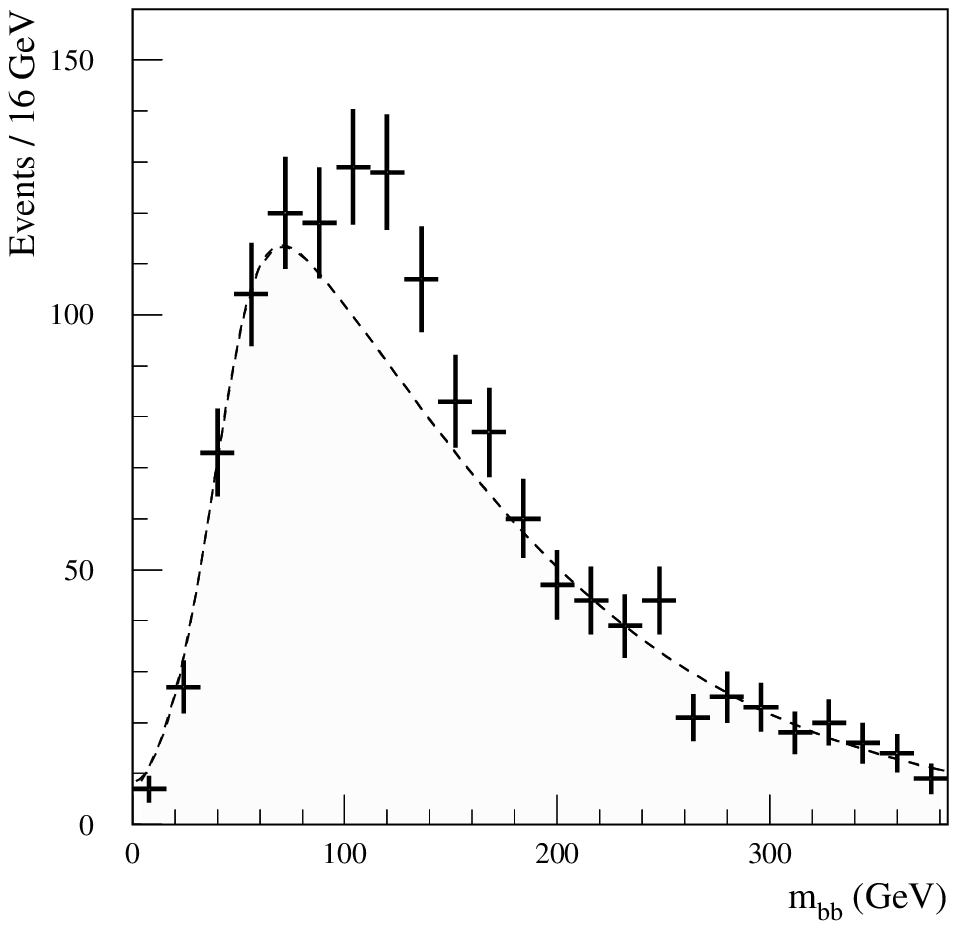}}
\caption[]{Invariant mass for the $b {\overline b}$ pair from
the Higgs decay in the process $pp\rightarrow t {\overline t} h$ at 
the LHC with $M_h=120~GeV$ (data points).
The dashed line is the background.  The error bars correspond
to the statistical uncertainty with an integrated luminosity
of $L=100~fb^{-1}$.\cite{lhctop}}
\label{fig:lhcfig}
\end{figure}

At the LHC, the process
 $pp\rightarrow t {\overline t} h$ has a significant
rate for $M_h< 130~GeV$ and we can 
look for $h\rightarrow b {\overline b}$, giving
the final state of $W^+W^- b {\overline b}
b {\overline b}$.  This is an important discovery channel for
a Higgs with $M_h< 130~GeV$.
The 
ATLAS experiment, with ${\cal L}=100~fb^{-1}$,
and assuming $M_h=120~GeV$ 
can measure ${\delta g_{tth}\over g_{tth}}\sim 16\%$. This
is illustrated in Fig. \ref{fig:lhcfig}.\cite{lhctop} 
The signal and the background have similar shapes at the LHC,
increasing the difficulty of this measurement.

\section{
Heavy Top as Inspiration for Model Builders}

We turn now to a discussion of the top quark in models
where there is new physics involving the top quark
beyond that of the Standard Model. 
The fact that the top quark is much heavier than the other
quarks gives it a special role in
electroweak symmetry breaking.  In this section, we examine the role
of the top quark in the minimal supersymmetric model and in models
with dynamical symmetry breaking.   We pay particular attention to how
data from the Tevatron and the LHC can help us to test these ideas.

The large value of the top quark mass has important consequences in
a supersymmetric (SUSY) model.\cite{mssm}  In the
SM, the Higgs boson mass is a free parameter, while in a SUSY model there
is an upper limit on the Higgs mass.  At tree level in 
the minimal SUSY model, $M_h< M_Z$, but
the large top quark mass, and the subsequent large coupling to the Higgs
boson,
\begin{equation}
{m_t \sqrt{2}\over v}\sim 1,
\end{equation}
 lead to large radiative corrections
to the prediction for the Higgs boson mass, raising the lower  limit
on the Higgs mass significantly.
  In addition, corrections due to the large top quark mass typically imply
that the scalar partner of the top, the stop, is the lightest
squark.

A further difference between the SM and a SUSY model which is
 due to the large
top quark mass is the understanding of the mechanism of electroweak 
symmetry breaking in a SUSY model.
 If the supersymmetric model is embedded in a grand
unified model, then the simplest version (mSUGRA)
 has the interesting
property that the evolution of the parameters
from the GUT scale to the weak scale 
 drives the Higgs mass-squared negative as a consequence of the
large value of $M_t$, 
and therefore
explains electroweak symmetry breaking.  In fact, this 
mechanism ${\it requires}$
 that the top quark have a mass near its observed value.

In models with dynamical symmetry breaking, the large value of the top
quark mass can imply that the top quark is different from the lighter quarks, 
perhaps having different gauge interactions
in an extended gauge sector.
  In Section 5.6,
 we discuss topcolor
as an example of a model of dynamical symmetry breaking where the top quark 
plays a special role.  Many possibilities for dynamical symmetry breaking
in electroweak interactions are detailed in the review by Hill and 
Simmons.\cite{dynsym}

\subsection{EWSB and the top quark mass}

We begin by considering the importance of the top quark mass in the SM. 
In the SM, electroweak symmetry breaking occurs through the
interactions of an $SU(2)_L$ doublet scalar field, $\Phi$, with
a scalar potential,\cite{hhg}
\begin{figure}[t]
\centerline{{\epsfysize=3.in\epsffile[0 0 600 600]{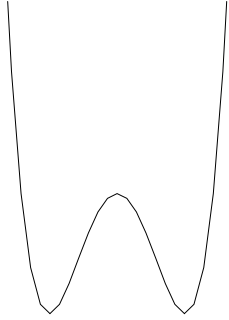}}}
\caption[]{Higgs potential in the Standard Model with $\mu^2<0$.}
\label{higgsfig}
\end{figure}
\begin{equation}
V=\mu^2\Phi \Phi^\dagger+\lambda(\Phi \Phi^\dagger)^2~~.
\label{potential}
\end{equation}
If $\mu^2<0$, the neutral component of the Higgs doublet
gets a vacuum expectation value,
\begin{equation}
\langle \Phi_0\rangle ={v\over\sqrt{2}}
\end{equation}
 and the scalar potential has the 
shape shown in Fig. \ref{higgsfig}. 
Electroweak symmetry breaking (EWSB) occurs and
masses are generated for the $W$ and $Z$ gauge bosons.
After EWSB, there remains in the 
spectrum a physical
particle, the Higgs boson, $h$, and
at tree level  the quartic self-
coupling of the Higgs boson, $\lambda$, is related to the Higgs mass by
\begin{equation}
M_h^2=2\lambda v^2 .
\end{equation}
A measurement of this relationship would test the mechanism of EWSB.

Eq. \ref{potential} is the scalar
potential at the electroweak scale, $\Lambda \sim v$.  All parameters
of the electroweak theory
 change
with the  energy scale $\Lambda$ in a precisely calculable way,
and so will be different at higher energies.
The most interesting parameters for us are 
the Higgs quartic self-coupling, $\lambda$, and the top quark
Yukawa coupling, $g_{tth}={M_t\over v}$.  To  one loop accuracy,\footnote{
$g$ and $g^\prime$ are the $SU(2)_L$ and $U(1)_Y$ gauge coupling 
constants.}
\begin{eqnarray}
\Lambda{d \lambda\over d \Lambda}&=&
{1\over 16\pi^2}\biggl[24 \lambda^2-(A-12g_{tth}^2)\lambda+B-12
 g_{tth}^4\biggr]
\nonumber \\ 
\Lambda{d g_{tth}\over d \Lambda}&=&
{1\over 16 \pi^2}\biggl({g_{tth}\over 12}\biggr)
\biggl[54 g_{tth}^2-17 g^{\prime 2}-27 g^{\prime 2} -96g_s^2]
\end{eqnarray}
where,
\begin{eqnarray}
A&=&3 (3 g^2+g^{\prime 2} )\nonumber \\
B&=&{3\over 8}
\biggl(2 g^4+(g^2+g^{\prime 2})^2 \biggr).
\label{rges}
\end{eqnarray}
We have omitted the equations for the running of the gauge
coupling constants.  They can be found in Ref. \refcite{hhg}.

In order for electroweak symmetry breaking to occur, the quartic coupling
must remain finite at all scales,
\begin{equation}
{1\over \lambda(\Lambda)}>0.
\end{equation}
As $\lambda$ is evolved from the weak scale to higher energy,
it grows and eventually becomes infinite at some large scale. 
When $\lambda$ becomes large, the scaling of Eq. \ref{rges} can
be approximated,
\begin{equation}
\Lambda{d \lambda\over d \Lambda}={3\over 2 \pi^2}\lambda^2 ~.
\label{largel}
\end{equation}
Eq. \ref{largel} is easily solved,
\begin{equation}
{1\over \lambda(\Lambda)}={1\over \lambda(M_h)}
-{3\over 2 \pi^2} \log\biggl({\Lambda\over M_h}
\biggr)~.
\end{equation}
 The
position where
$\lambda(\Lambda)\rightarrow \infty$ is known as the Landau Pole.
Since $\lambda$ is related to the Higgs mass,
$M_h^2=2 v^2\lambda$, the requirement
that the theory not approach the Landau pole gives an upper bound
on the Higgs boson mass,
\begin{equation}
M_h^2<{4 \pi^2 v^2\over 3 \log\biggl(\Lambda/M_h
\biggr)} ~~.
\end{equation}

\begin{figure}[t,b]
\begin{center}
{{\epsfysize=4.5in\epsffile
[ 0 0 600  600]{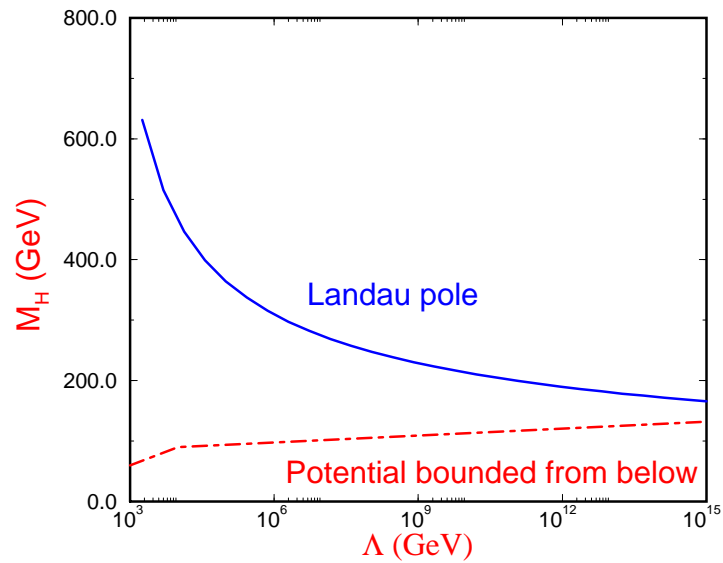}}} 
\caption[]{Limits on the Higgs boson mass as a function of 
the cutoff scale $\Lambda$.  The theoretically consistent region
lies between the two curves.\cite{sher}  }
\label{higgslim_fig}
\end{center}
\end{figure}

This means that at the scale $\Lambda$ there must be some
new physics beyond that of the SM which enters.  The SM simply
makes no sense at energy
scales larger than $\Lambda$ since in this region the
Higgs quartic self-coupling is infinite.
Assuming no new physics before the GUT scale,
$\Lambda\sim 10^{16}~GeV$, gives a bound on
the Higgs mass of\cite{sher,quiros} 
\begin{equation}
M_h< 160~GeV ~.
\end{equation}
This  bound corresponds to the upper curve in Fig. \ref{higgslim_fig}.
Non-perturbative studies using lattice gauge theory confirm the 
validity of this limit.\cite{lattice}

Conversely, we require that the potential be bounded from 
below, which is equivalent to the restriction  that
$\lambda>0$ at large $\Phi$.  The scaling of the quartic coupling 
near $\lambda=0$ is,
\begin{equation}
\Lambda{d \lambda\over d \Lambda}=
{1\over 16\pi^2}\biggl[B-12
 g_{tth}^4\biggr]
\end{equation}
which can be solved,
\begin{equation}
\lambda(\Lambda)=\lambda(M_h)+{B-12g_{tth}^4\over 16\pi^2}
\log\biggl({\Lambda\over M_h}\biggr)~.
\end{equation}
The requirement that $\lambda(\Lambda)>0$ gives
 a lower bound on the Higgs
mass,
\begin{equation}
{M_h^2\over 2 v^2}>{12 g_{tth}^4-B\over 16 \pi^2}\log\biggl({\Lambda
\over M_h}\biggr)~~.
\end{equation}
A complete approach must include the full set of renormalization group 
equations, including the running of the coupling constants for
large values of $\Phi$.

The restrictions on the Higgs mass
from consistency of the theory
 are shown in Fig. \ref{higgslim_fig}, where the 
allowed region for a scale $\Lambda$ is the area between the
curves, known as the ``chimney''.
 At a given scale $\Lambda$,
the Higgs mass  has a lower limit  from bounding
the potential from below  and an upper limit
from the requirement that Higgs quartic coupling, $\lambda$,  
be finite. 
The exact shape of the
allowed region  depends  sensitively on the value 
of  $M_t$.\cite{sher,quiros}

The SM with a Higgs mass in the $130~GeV$ region is consistent at
energy scales all the way to the Planck scale.  The theory is still
unsatisfactory however, because it
provides no explanation of why $\mu^2<0$, as required for EWSB.  In
addition, the  
Higgs mass is a free parameter in the  SM and there is no understanding
of why it should have  any particular value.
In the next section, we will see that low energy supersymmetry can
potentially
solve both these problems.

\subsection{Supersymmetry and the Top Quark}

TeV scale supersymmetric models explain several mysteries about the SM:
\begin{itemize}
\item
A heavy top  quark and mSUGRA (minimal supergravity) explain  why  $\mu^2<0$.
\item
A heavy top quark and the MSSM (minimal supersymmetric Standard
Model) explain why 
the Higgs mass may light, as suggested by precision measurements.
\end{itemize}
In this section, we give a brief review of phenomenological models
of supersymmetry and
discuss the relationship between the top quark mass and the requirement
that supersymmetry exist below the TeV scale. 

Supersymmetry is a symmetry which relates
the masses and couplings of  particles of differing
spin.  To each SM chiral fermion is associated a complex scalar and each of
the $SU(3)\times SU(2)_L\times U(1)_Y$ gauge bosons similarly
 acquires a Majorana fermion
partner. 
Supersymmetry connects particles of differing spin, but
with all other
characteristics the same.\cite{wessbagger} 
In an unbroken supersymmetric model, particles and their superpartners
have identical masses.  It is clear, then, that {\bf
supersymmetry must be a broken symmetry} if it is to be a theory
of low energy interactions.
There is no scalar particle, for example, with the mass and
quantum numbers of the electron.  In fact, there are no candidate
supersymmetric scalar partners for any of the fermions in the
experimentally observed 
spectrum.  The most general soft breaking of supersymmetry introduces
masses and mixing parameters and the theory loses much of its capabilities
for specific predictions.

The Higgs sector in a SUSY model  always contains at least two
$SU(2)_L$ doublets.
 In the supersymmetric extension of
the Standard Model, the Higgs doublet acquires a SUSY
partner which is an $SU(2)_L$ doublet of
Majorana  fermion fields, 
 ${\tilde h}_1$  (the Higgsinos), which   
 contribute to the triangle $SU(2)_L$ and $U(1)_Y$ gauge anomalies as
described in Section 2.1.
Since the fermions of the Standard Model have exactly the
right quantum numbers to cancel these anomalies, it follows
that the contribution from the fermionic partner of the
Higgs doublet remains uncancelled.\cite{peskin,qu}    
This contribution must be
 cancelled by adding a second Higgs doublet (and its Majorana
fermion partner) with 
precisely the opposite $U(1)_Y$ quantum numbers from the
first Higgs doublet.

All supersymmetric models  thus have at least
 2-Higgs doublets
and so  the minimal supersymmetric model has
5 physical Higgs bosons:  $h^0,H^0,A^0,H^\pm$.  The ratio of the
vacuum expectation values of the neutral Higgs bosons,
\begin{equation}
\tan\beta\equiv {v_2\over v_1}
\end{equation}
is a fundamental parameter of the theory.
At tree level, the neutral Higgs boson masses are predicted
in terms of two parameters, which are usually taken to be
$\tan\beta$ and the mass of the pseudoscalar Higgs boson, $M_A$.
The neutral Higgs boson masses are then predicted,
\begin{equation}
M_{h,H}^2={1\over 2} 
 (M_A^2+M_Z^2) \mp
{1\over 2}
\sqrt{(M_A^2+M_Z^2)^2-  4 M_Z^2 M_A^2\cos^2 2 \beta}
\end{equation}
which gives an
upper bound on the lightest Higgs mass: 
\begin{equation}
M_h < M_Z \mid \cos 2 \beta\mid < M_Z
\label{mhsusy}
\end{equation}
which is excluded experimentally.  (In a SUSY model, the LEP and LEPII 
results require $M_h>89.8~GeV$.\cite{pdg})

\begin{figure}[t]
\begin{center}
{{\epsfysize=4.in\epsffile
[ 0 0 600  600]{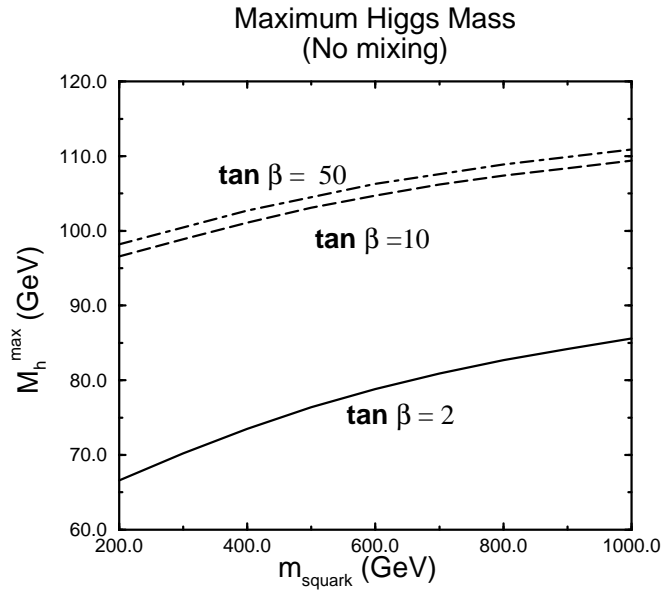}}} 
\caption[]{Upper bound on Higgs mass in MSSM at one loop and
with no soft trilinear mixing.\cite{twoloop}}
\label{fig:upp}
\end{center}
\end{figure}

The Higgs mass prediction in a supersymmetric model  
receives large radiative corrections from top  and stop squark loops.
At 1-loop, and neglecting possible soft SUSY breaking tri-linear
scalar mixing, the lightest Higgs mass becomes:
\begin{eqnarray}
M_{h}^2& =&
{M_A^2+M_Z^2+\epsilon\over 2} 
-{1\over 2}  \biggl\{(M_A^2+M_Z^2+\epsilon)^2-
  4 M_Z^2 M_A^2\cos^2 2 \beta  
\nonumber \\ && 
-4 \epsilon (M_A^2\sin^2\beta +M_Z^2 \cos^2\beta)\biggr\}^{1/2}
\end{eqnarray}
where,
\begin{eqnarray}
\epsilon = &&{3 G_F\over \sqrt{2}\pi^2}{M_t^4
\over \sin^2\beta}\log\biggl({{\tilde M}_t^2\over
M_t^2}\biggr) ,
\label{mhcor}
\end{eqnarray}
and ${\tilde M}_t$ is the average stop squark mass.
For unbroken supersymmetry, the top and stop have the same masses, 
$M_t={\tilde M}_t$, and the bound of Eq. \ref{mhsusy} is recovered.  
The mass splitting between the top and stop in a broken SUSY theory
changes the
upper bound on  the Higgs mass to,
\begin{equation}
M_h^2 < M_Z^2 \cos^2 2 \beta +\epsilon \sin^2\beta .
\end{equation}
The large $M_t^4$ corrections raise the Higgs mass
bound above the experimental
limit, as
shown in Fig. \ref{fig:upp}.

There are many analyses\cite{twoloop}
which include a variety of two-loop effects, renormalization
group effects, etc., but the important point is that for
 given values  of $\tan\beta$, the scalar mixing
parameters, and the squark masses,
 there is an upper bound
on the lightest neutral Higgs boson mass.  
  For large values of $\tan\beta$
the limit is relatively insensitive to the value of
$\tan\beta$ and with  a squark mass less than about $1~TeV$,
the upper limit on the Higgs mass is about $110~GeV$ if mixing in
the top squark sector is negligible.
For large mixing, this limit is raised to
around $130~GeV$.

\subsection{mSUGRA}
The minimal supersymmetric model (MSSM) is constructed by adding to the
Lagrangian  all of
the soft supersymmetry breaking terms allowed by the 
gauge symmetries.  These include masses for all the
new SUSY particles and tri-linear scalar mixing terms.
 The most general Lagrangian of this
type has 105 new parameters beyond those
of the Standard Model.  In order to make definite predictions, the
simplifying assumption that the 
 soft parameters unify at $M_{GUT}$ is often made.  (The resulting model
is usually called mSUGRA or the CMSSM).
In this framework the
model  is  completely specified by five parameters:

\begin{itemize}
\item
Common scalar mass at  $M_{GUT}$, ${\tilde m}_0$
\item
Common gaugino mass at  $M_{GUT}$, $M_{1/2}$
\item
1 Higgs mass-squared, $m_{12}^2$
\item
1 tri-linear coupling at  $M_{GUT}$, $A_0 \lambda_F$
\item
sign$({\hat\mu})$
\end{itemize}
The parameter ${\hat\mu}$ describes the mixing between the two Higgs doublets.
At the GUT scale, the Higgs masses are,
\begin{equation}
M_{h,H}^2(M_{GUT})=m_0^2+{\hat\mu}^2,
\end{equation}
while the squark  and slepton masses are
\begin{equation}
M_{\tilde q}^2(M_{GUT})=M_{\tilde l}^2(M_{GUT})=m_0^2~.
\end{equation}
The theory must explain why the lightest
neutral Higgs mass-squared becomes negative
at the weak scale 
and breaks the electroweak symmetry, while the scalar fermion masses
remain positive so as to not break color and electromagnetism.

In an mSUGRA model, the
 masses and parameters are evolved using the renormalization 
group equations (RGE) to scale from $M_{GUT}$ down to the weak scale in order to make
predictions for the masses at the weak scale. 
The large top quark mass plays an important role here.
 We can understand the basic
scaling by 
considering only  the Higgs doublet, $H_2$,\footnote{$H_2$ is
the $SU(2)_L$ doublet with isospin $Y= +{1\over 2}$.}
the scalar partners of the third generation $SU(2)_L$
fermion doublet, $({\tilde t}_L,{\tilde b}_L)\equiv
 {\tilde Q}_3$, and
the scalar partner of the right-handed top quark,  ${\tilde t}_R$.
The 
renormalization group scaling is given by:\cite{maha}
\begin{eqnarray}
{dM_{H_2}^2\over d\log(\Lambda)}
&=& {1\over 8 \pi^2} 
(3 g_{tth}^2 X_t-3 g^2 M_2^2-g^{\prime 2} M_1^2)\nonumber \\
& & \nonumber \\
{d{\tilde M}_{ t_R}^2\over d\log(\Lambda)}
&=& {1\over 8 \pi^2} (2 g_{tth}^2 X_t-
{16\over 3} g_s^2 M_3^2-{16\over 9}g^{\prime 2} M_1^2)\nonumber \\
&& \nonumber \\
{d{\tilde M}_{Q_3}^2\over d\log(\Lambda)}
&=& {1\over 8 \pi^2} ( g_{tth}^2 X_t-{16\over 3} g_s^2 M_3^2
-3 g^2M_2^2
-{1\over 9} g^{\prime 2} M_1^2)\nonumber 
\end{eqnarray}
where,
\begin{eqnarray}
X_t&\equiv& M^2_{H_2}+{\tilde M}^2_{Q_3}+{\tilde M}^2_{{ t_R}}
+A_t^2 ~.
\end{eqnarray}
$M_{3,2,1}$ are the gaugino masses associated with
the $SU(3)\times
SU(2)_L\times U(1)_Y$ gauge sector, $g_{tth}$ is the coupling of
the Higgs to the top quark in the MSSM, and $A_t$ is the soft SUSY
mixing term in the top sector. 
The
QCD term   (proportional to $g_s^2$) makes ${\tilde M}_{t_R}^2,
{\tilde M}_{Q_3}^2$ increase faster than $M_{H_2}$ as the
scale is reduced from $M_{GUT}$.
The
Yukawa term  (proportional to $g_{tth}$) 
drives $M_{H_2}^2 < 0$, while the other scalar
mass-squared terms remain positive.  This behavior is a 
consequence of the large $M_t$ value.

Neglecting the gauge couplings, we have:
\begin{equation}
{d\over d\log(\Lambda)}
\left(
\begin{matrix}
M_{H_2}^2\cr
{\tilde M}_{tR}^2\cr
{\tilde M}_{Q_3}^2\cr
\end{matrix}
\right)=
{g_{tth}^2\over 8 \pi^2} X_t
\left(
\begin{matrix}
3\cr
2\cr
1\cr
\end{matrix}
\right)~,
\end{equation}
which can be easily solved,
\begin{equation}
M_h^2(\Lambda)=M_h^2(M_{GUT})-{3g_{tth}^2\over
8\pi^2}X_t\log({M_{GUT}\over \Lambda})~.
\end{equation}
For heavy $M_t$, $M_h^2$  becomes negative and triggers EWSB.  It is
interesting that this behavior ${\it requires}$ $M_t\sim 175~GeV$.\cite{barger}

\begin{figure}[t,b]
{\centerline{\epsfysize=3.0in\epsffile[32 46 593 406]
{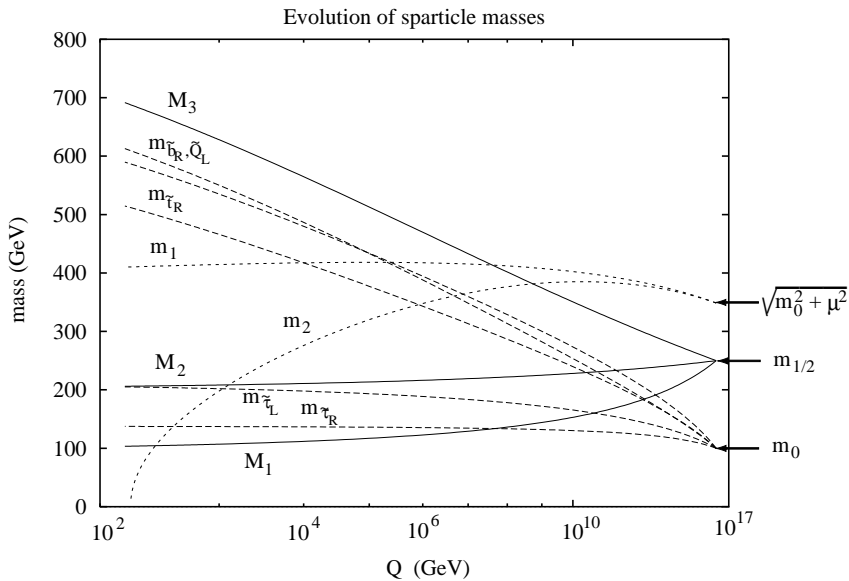}}} 
\caption[]{Typical pattern of masses in an 
mSUGRA model.\cite{barger}.}
\label{fig:mass_fig_susy}
\end{figure}

The RGE scaling from the GUT scale to the
weak scale in  mSUGRA type models yields a
a typical pattern of masses, as seen in Fig. \ref{fig:mass_fig_susy}.
  The gluino is the
heaviest gaugino, while the squarks are significantly heavier than
the sleptons.  

\subsection{Charged Higgs Decays}
The pattern of top  quark decays is changed in the MSSM from that of
the SM, due to the extended Higgs sector.
  The presence of a charged Higgs boson allows the 
decay,
$t\rightarrow H^+ b$ if the mass of the charged Higgs
is less than $M_t-m_b$.  This decay looks similar to $t\rightarrow W b$ 
and can have a significant rate.
The coupling of the charged Higgs to the $t~b$ system is,
\begin{equation}
g_{tH^+b} \sim m_t \cot\beta P_+ +m_b\tan\beta P_-
\end{equation}
and so this decay is relevant both for very large
and for very small $\tan\beta$.
For small $\tan\beta$, $H^+\rightarrow c s$ and $H^+\rightarrow t^*b$,
while for
large $\tan\beta$, $H^+\rightarrow \tau\nu$.  These decays would suppress the
rate for top decays to leptons plus jets 
in the SM top search. In Run I, the
Tevatron experiments placed a limit
on $M_{H^+}$ of roughly $M_{H^+}>100~GeV$ for $\tan\beta<1$ or 
$\tan\beta >50$.\cite{charged}

\subsection{The Top squark}
The mass-squared matrix for the scalar partners of the left and 
right handed top quarks (${\tilde t}_L,{\tilde t}_R$) includes
mixing between
the partners of  the left- and right- handed top:
$$
{\tilde M}_t^2=
\left(
\begin{matrix}
{\tilde M}_{Q_3}^2+M_t^2 
& M_t(A_t-{\hat\mu} \cot\beta)\cr
~~~
+M_Z^2 L_t\cos 2 \beta & \cr  
& {\tilde M}_{t_R}^2+M_t^2
\cr
M_t(A_t-{\hat\mu}\cot\beta)& ~~~
+M_Z^2 R_t \cos 2 \beta
\cr
\end{matrix}\right)
$$
The mixing is proportional to the top quark mass and so the large value
of $M_t$ can cause one of the top squarks to be relatively light.  If
the scalar mass scale,
${\tilde M}_{Q_3}$ is much larger than $M_Z, M_t, A_t$, then
the stop masses are
 roughly degenerate $\sim {\tilde M}_{Q_3}$.
On the other hand if 
${\tilde M}_{Q_3}\sim M_Z, M_t, A_t$,
then the
large mixing effects 
drive the stop mass to a small value and the stop squark becomes
the lightest squark.  

It is possible to search for a light stop in the decays of the
top,
$t\rightarrow {\tilde t} {\tilde \chi}^0$,
where ${\tilde \chi}^0$ is the lightest neutral
SUSY particle.  The limits then
depend on the assumed branching ratio for the top into stop decays.  The
Run I data was sensitive to stop and chargino masses in the $100~GeV$
range.\cite{topsquark_lim}

\subsection{The Top Quark  and Dynamical Symmetry Breaking}

The top quark has a special role in technicolor models where  a $t {\overline t}$
condensate can play the part of the Higgs boson in generating EWSB. 
This can happen in models where there is a new strong interaction felt by
the top quark, but not by the lighter quarks.
  The proto-type model of this type is called ${\it topcolor}$.\cite{topref}
 A review of this class of models can be
found in Ref. \cite{dynsym}.
Topcolor models are constructed by expanding the strong gauge group
from the $SU(3)$ of QCD to,
\begin{equation}
SU(3)_{tc}\times SU(3)\rightarrow SU(3)_{QCD}
\end{equation}
$SU(3)_{tc}$ couples only to the 3rd generation with
a coupling $g_{tc}$,
while the second
 $SU(3)$ gauge symmetry couples only  to the
first two generations.  
The symmetry is  broken
 to $SU(3)_{QCD}$ at a scale $M$.  Below $M$, the
     gauge bosons  of the $SU(3)_{tc}$ group, the ``topgluons'', 
are massive and lead to
effective 4-Fermi interactions between the top quarks:
\begin{equation}
{\cal L}\sim {g_{tc}^2\over M^2}
({\overline Q}_{3L} \gamma_\mu {T^A\over 2}
Q_{3L})({\overline t}_R\gamma^\mu {T^A\over 2}t_R),
\end{equation}
where $T^A$ is the $SU(3)$ generator in the adjoint representation
and $Q_{3L}$ is the left-handed $(t_L,b_L)$ doublet.
If the $SU(3)_{tc}$ coupling, $g_{tc}$, is larger than a critical value, a top
condensate forms, $\langle {\overline t} t\rangle $,
 which breaks the electroweak symmetry and generates
a mass for the top quark.
One must arrange that $\langle {\overline t}
  t \rangle  \ne 0$
and $\langle {\overline b}  b\rangle=0$.  This is usually done
by adding an additional $U(1)_h$ gauge symmetry under which the $t$
and $b$ quarks transform differently. 

This class of models can be tested at the Tevatron and the LHC by
searching for effects of the top gluons and the new $Z^\prime$, and
through $B$ decays.\cite{simmons_k}  Since the $Z^\prime$ 
boson corresponding to the $U(1)_h$ and the top gluons
of $SU(3)_{tc}$  couple differently
to the third generation and  to 
the lighter quarks, they lead to flavor changing
neutral currents. Present limits from $B$ meson mixing require
that the topgluon be heavier than $3-5~TeV$ and $M_{Z^{\prime}}>
7-10~GeV$.\cite{btop}

\section{Conclusions}
The story of the top quark is far from over.  Its properties and its
role in the SM and in physics beyond the SM will be further elucidated
in the emerging Run II data at the Tevatron.  The increased statistics
of Run II will enable more precise measurements of the mass, couplings,
and decay properties of the top. The LHC will provide 
even  more insight into the role of the top quark.  
It is possible that surprises are just around the corner!

\section*{Acknowledgements}
I would like to thank K.T. Mohanthapa,
 H. Haber, and A. Nelson for organizing such an enjoyable
and successful school.

\end{document}